\documentclass[a4paper,reqno,12pt]{amsart}
\usepackage[margin=2.5cm,papersize={19.75cm,28cm}]{geometry}

\usepackage{hyperref}
\usepackage{amssymb}
\usepackage{eucal}
\usepackage{eufrak}
\usepackage{graphics}
\usepackage{graphicx}
\usepackage{epsfig}
\usepackage{multicol}
\usepackage{xypic}
\usepackage{amsmath}
\usepackage{wasysym}
\usepackage{dsfont}
\usepackage{xypic}
\usepackage{amsmath}
\usepackage{color}

\newtheorem{theorem}{Theorem}[section]

\newtheorem{definition}{Definition}[section]

\newtheorem{corollary}[definition]{Corollary}
\newtheorem{remarkth}[definition]{Remark}
\newenvironment{remark}{\begin{remarkth}\upshape}{\end{remarkth}}

\newcommand{\se}{\mathfrak{se}(3)}

\newcommand{\Flder}{\rightarrow}
\newcommand{\lp}{\left(}
\newcommand{\rp}{\right)}
\newcommand{\lc}{\left\{}
\newcommand{\rc}{\right\}}
\newcommand{\der}{\partial}

\newcommand{\ra}{\rightarrow}

\newcommand{\bra}{\langle}
\newcommand{\ket}{\rangle}
\newcommand{\R}{\mathds{R}}      

\newcommand{\lcf}{\lbrack\! \lbrack}
\newcommand{\rcf}{\rbrack\! \rbrack}

\usepackage{amssymb}

\newcommand{\proa}{A^*G \mbox{$\;$}_{\tau^*} \kern-3pt\times_\alpha
G \mbox{$\;$}_\beta \kern-3pt\times_{\tau^*} A^*G}

\newcommand{\alg}{\mathfrak{so}(3)}

\newcommand{\e}{\mbox{exp}}
\newcommand{\Ad}{\mbox{Ad}}
\newcommand{\ca}{\mbox{cay}}
\newcommand{\ad}{\mbox{ad}}

\newcommand{\al}{\mathfrak{g}}
\newcommand{\dal}{\mathfrak{g}^{*}}

\def\lcf{\lbrack\! \lbrack}
\def\rcf{\rbrack\! \rbrack}

\begin{document}

\title[variational integrators for control systems with symmetries]{Variational integrators for underactuated mechanical control systems with
symmetries}
\author{Leonardo Colombo}
\address{L.Colombo: Department of Mathematics, University of Michigan. EH 3828, 530 Church Street, Ann Arbor, MI, 48109-1043. USA } \email{ljcolomb@umich.edu}

\author{Fernando Jim\'enez}
\address{F.Jim\'enez: Zentrum Mathematik der Technische Universit\"at M\"unchen, D-
85747 Garching bei M\"unchen, Germany} \email{fjimenez@ma.tum.de}

\author{David Mart\'in de Diego}
\address{D.\ Mart\'{\i}n de Diego: Instituto de Ciencias Matem\'aticas (CSIC-UAM-UC3M-UCM), Calle Nicol\'as Cabrera 15, 28049 Madrid, Spain} \email{david.martin@icmat.es}

\thanks{This work has been  supported by  MEC (Spain) Grants MTM
2010-21186-C02-01, MTM2009-08166-E,  MINECO: ICMAT Severo Ochoa
project SEV-2011-0087 and IRSES-project Geomech-246981. LC wants to
thank CSIC for a JAE-Pre grant. The research of FJ has been
supported in its final stage by the DFG Collaborative Research
Center TRR 109, ``Discretization in Geometry and Dynamics''. The authors wish to thank to the referees by the
relevant and important comments that have been improved this work in
its final version.}



\maketitle

\begin{abstract}
Optimal control problems for underactuated
mechanical systems can be seen as a higher-order variational problem
subject to higher-order constraints (that is, when the Lagrangian
function and the constraints depend on higher-order derivatives such
as the acceleration, jerk or jounces). In this paper we discuss the
variational formalism for the class of underactuated mechanical
control systems when the configuration space is a trivial principal
bundle and the construction of variational integrators for such
mechanical control systems.

An interesting family of geometric integrators can be defined using
discretizations of the Hamilton's principle of critical action. This
family of geometric integrators is called variational integrators,
being one of their main properties the preservation of geometric
features as the symplecticity, momentum preservation and good
behavior of the energy. We construct variational
integrators for higher-order mechanical systems on trivial principal
bundles and their extension for higher-order constrained systems and
we devote special attention to the particular case of underactuated
mechanical systems
\end{abstract}



\section{Introduction}

The construction of variational integrators have received a lot of
interest in the recent years from the theoretical and applied points
of view. The goal of this paper is to develop variational
integrators for optimal control problems of mechanical systems
defined on a trivial principal bundle, paying particular attention
to underactuated mechanical control systems. Underactuated control
systems are characterized by the fact that they have more degrees of
freedom than actuators.

The presence of underactuated mechanical systems is ubiquitous in
engineering applications; for instance, the
underactuation may arise from a failure in the fully actuated regime
or in the design of less costly devices. Optimal control problems
of underactuated mechanical systems can be seen as a variational
problem involving Lagrangians defined on higher-order tangent
bundles subject to higher-order constraints. That
is, Lagrangian and constraint functions depending not just on
positions and velocities, but also on acceleration and sometimes
higher-order derivatives of the curve in the configuration space.
The purpose of the optimal control problem is to find a curve of the
state variables and control inputs which satisfies the controlled
equations and minimizes a cost function subject to initial and final
boundary conditions. We will use the equivalence
between optimal control problems of underactuated mechanical systems
and second-order variational problems with second-order constraints
(see \cite{Bl}) when the configuration manifold is a trivial
principal bundle. To develop this process it is necessary to obtain
the second-order Euler-Lagrange equations, calculation which is
showed in great detail. Furthermore, the Lagrange-Poincar\'e
equations follow from the Euler-Lagrange ones after applying a
symmetry reduction procedure. Moreover, recent
developments in the theory of $k-$splines on $SO(3)$ and the
Clebsh-Pontryagin optimal control problem (see
\cite{GBHR10,FHR2010}), suggest that the higher-order situation may
be relevant. Consequently, we extend the previous calculation to
this case, obtaining Lagrange-Poincar\'e equations using a left
trivialization of the tangent bundle of a Lie group.

Higher-order variational problems also have been
used in the applications to longitudinal studies in computational
anatomy. This kind of studies seeks, among other goals, to determine
a path that interpolates optimally through a time-order series of
images or shapes. Depending on the specific application, the
interpolant will be require to have a certain degree of
spatiotemporal smoothness. If a higher-order degree of smoothness is
required, a natural approach is to investigate higher-order
variational formulation of the interpolation problem (see for
example \cite{BHM},\cite{alexyjair}, \cite{FHR2010} and
\cite{meier}).

In order to obtain variational integrators for the systems described
above, we extend the theory of discrete mechanics, which is based on
discrete calculus of variations, to higher-order systems subject to
higher-order constraints. The discrete approach proposed in this
work (we will employ {\it our setting} or {\it our approach}
henceforth for sake of simplicity) follows the ideas proposed by
Marsden and Wendlandt \cite{welant} and Marsden and West
\cite{mawest} for first order systems without constraints. In
particular, we shall develop the discrete analogue of higher-order
Euler-Lagrange and Lagrange-Poincar\'e equations (introduced by
Gay-Balmaz, Holm and Ratiu in \cite{GHR2011}) for trivial principal
bundles and its extension to constrained systems. With this
particular purpose, we use the discrete Hamilton's principle and the
adding of Lagrange multipliers in order to obtain discrete paths
that approximately satisfy the dynamics and the constraints. Such
formulation gives us the preservation of important geometric
properties of the mechanical system, such as momentum,
symplecticity, group structure and good behavior of the energy (see
\cite{Hair}). The preservation of these geometric
properties produces an improved qualitative behavior, and also a
more accurate long-time integration than with a standard
integrator.

The methods studied in this paper are founded on
recently developed structure-preserving numeric integrators for
optimal control problems (see
\cite{Bl,BHM,CoJiMdD,CoMaZu210,CoMaZu2011,JiMa,Marin,KM,TL,MMS,objuma}
and references therein) based on solving a discrete optimal control
problem as a discrete variational problem with constraints (see
\cite{Bl,CoMaZu210,CoMaZu2011} for the continuous counterpart). These numerical integrators are used for simulating
and controlling the dynamics of satellites, spacecrafts, underwater
vehicles, mobile robots, helicopters and wheeled vehicles
\cite{BlochLeok,bullolewis,LeeLeok}. More concretely, previous
results (upon which the present ones are based) can be found in
\cite{CoJiMdD} and \cite{JKM}, where mainly the treated
configuration manifolds are Lie groups and in the second work the
authors solve optimal control problems with a different view point,
where they construct variational integrators for systems with
piecewise controls. Previous numerical methods are
developed only for the case of higher-order tangent bundles
\cite{CoMaZu2011},\cite{CoMdDZu2013} or several copies of the Lie
algebra of a Lie group (see for example \cite{BHM}, \cite{CoJiMdD}).
Both correspond to numerical methods for higher-order Euler-Lagrange
equations and Euler-Poincar\'e equations, respectively. However,
real systems are typically modeled over a manifold which admits a
Lie group of symmetries (mainly $SE(2)$, $SO(2)$, $SE(2)$ and
$SE(3)$). Therefore, applying standard reduction theory we derive a
system defined on a quotient bundle. This bundle is, in several
cases, the cartesian product of the previously mentioned spaces;
that is, a higher-order tangent bundle and several copies of a Lie
algebra associated with the symmetry Lie group. Thus we design
geometric integrators for this type of spaces, which includes the
previous ones as particular cases, and we apply this construction to
the desing of variational integrator for optimal control problems of
mechanical systems in presence of Lie groups symmetries (see
\cite{Borda},\cite{borda2}, \cite{CoMdDZu2013}, \cite{CoMaZu2011}).

To be self-contained, we introduce a brief background in discrete
mechanics.




\subsection{General Background}
A \textit{discrete Lagrangian} is a map $L_d\colon Q \times Q\to
\R$, which may be considered as an approximation of the integral
action defined by a continuous  Lagrangian $L\colon TQ\to \R,$
\[
L_d(q_0, q_1)\approx \int^h_0 L(q(t), \dot{q}(t))\; dt,
\]
where $q(t)$ is the unique solution of the Euler-Lagrange equations
for $L$; $q(0)=q_0$, $q(h)=q_1$ and the time step $h>0$ is small
enough.

Given the grid $\{t_{k}=kh\mid k=0,\ldots,N\}$,
with $Nh=T$, define the discrete path space
$\mathcal{C}_{d}(Q):=\{q_{d}:\{t_{k}\}_{k=0}^{N}\ra Q\}.$ This discrete path space is isomorphic to the
smooth product manifold which consists of $N+1$ copies of $Q$. The
discrete trajectory $q_{d}\in\mathcal{C}_{d}(Q)$ will be identified
with its image $q_{d}=\{q_{k}\}_{k=0}^{N}$ where
$q_{k}:=q_{d}(t_{k}).$

Define the \textit{action sum} $\mathcal{A}_d\colon
\mathcal{C}_{d}(Q)\to \R$, associated to $L_d$,
by summing the discrete Lagrangian on each adjacent pair
$$
{\mathcal{A}_d}(q_{d}):=\sum_{k=1}^{N-1}  L_d(q_{k-1}, q_{k}),
$$
where $q_k\in Q$ for $0\leq k\leq N.$ Note that the discrete
action inherits the smoothness of the discrete Lagrangian.

The discrete variational principle then requires that $\delta
{\mathcal{A}_d}=0$ where the variations are taken with respect to
each point $q_k,$ $1\leq k\leq N-1$ along the path, and the
resulting equations of motion (a system of difference equations),
given fixed endpoints $q_0$ and $q_N,$ are
\begin{equation}\label{discreteeq}
 D_1L_d( q_k, q_{k+1})+D_2L_d( q_{k-1}, q_{k})=0,
\end{equation} where $D_1$ and $D_2$ denote the derivative of the discrete Lagrangian
with respect to the first and second arguments, respectively. These
equations are usually called \textit{discrete Euler--Lagrange
equations}.

If the matrix $D_{12}L_d(q_k, q_{k+1})$ is regular, it is possible
to define a (local) discrete flow $ \Upsilon_{L_d}\colon Q\times
Q\to Q\times Q$, by $\Upsilon_{L_d}(q_{k-1}, q_k)=(q_k, q_{k+1})$
from (\ref{discreteeq}). This discrete flow
preserves the (pre-)symplectic form on $Q\times Q,$ $\omega_d$, i.e.
$\Upsilon_{L_d}^*\omega_d=\omega_d$ (see \cite{welant},
\cite{mawest} and references therein).

Given an action of a Lie group $G$ on $Q$, we can consider the
$G-$action on $Q\times Q$ $g\cdot (q_{k},q_{k+1}):=(g\cdot
q_{k},g\cdot q_{k+1})$.  Denoting by $\mathfrak{g}$ the Lie algebra
of $G$ we can define two discrete momentum maps
\begin{eqnarray*}J_{d}^{\pm}(q_{k},q_{k+1}):&&\mathfrak{g}\rightarrow\R\\
&&\xi\mapsto\langle\Theta_{L_{d}}^{\pm}(q_k,q_{k+1}),\xi_{Q\times Q}(q_{k},q_{k+1})\rangle
\end{eqnarray*} for $\xi\in\mathfrak{g}.$ Here $\xi_{Q\times Q}(q_{k},q_{k+1}):=(\xi_{Q}(q_k),\xi_{Q}(q_{k+1}))$ where \[\xi_{Q}(q)=\frac{d}{dt}\Big|_{t=0} (exp(t\xi)\cdot q)
\] denotes the fundamental vector field and
$\Theta^{+}_{L_d}(q_{k},q_{k+1}):=D_{2}L_{d}(q_k,q_{k+1})dq_{k+1},$
$\Theta^{-}_{L_d}(q_{k},q_{k+1}):=-D_{1}L_{d}(q_k,q_{k+1})dq_{k}$
are the discrete Poincar\'e-Cartan 1-forms on $Q\times Q.$ If the
Lagrangian is $G-$invariant then $J_{d}:=J_{d}^{+}=J_{d}^{-}$ and
$J_{d}\circ\Upsilon_{d}=J_{d}.$

\subsection{Goals, contributions and organization of the paper.}
The main goal of this work is to develop
variational integrators for optimal control problems of
underactuated mechanical systems defined on a trivial principal
bundle. This is achieved in section \S\ref{section6}, where the
process is showed in great detail for two examples. With that
purpose, other contributions are presented previously. Namely: in
\S\ref{section3} the continuous Euler-Lagrange (Lagrange-Poincar\'e)
equations for higher-order tangent bundles when the configuration
manifold is a trivial principal bundle are obtained, particularly in
theorem \ref{ContHOTPB}; section \S\ref{section4} is devoted to the
construction of variational integrators for higher-order mechanical
systems with symmetries where we obtain the discrete higher-order
Euler-Lagrange and Lagrange-Poincar\'e equations in theorem
\ref{DiscHOPB}; in \S\ref{section5} higher-order constraints are
added into the picture and consequently we obtain the equations of
motion for constrained system, both in the continuous and discrete
cases by using Lagrange multipliers; and, finally, in
\S\ref{section6} we study optimal control problems, we apply the
techniques developed previously in order to obtain the design of
geometric numerical integrators for this kind of optimal control
problems and explore two examples: the optimal control of a vehicle
whose configuration space is the Lie group $SE(2),$ and the
associated optimal control problem of a homogeneous ball rotating on
a plate.





\section{Higher-order Euler-Lagrange equations on trivial principal
bundles}\label{section3}


\subsection{Higher-order tangent bundles}

In this subsection we recall some basic facts of the higher-order
tangent bundle theory. At some point, we will particularize this
construction to the case when the configuration space is a Lie group
$G$.  For more details see \cite{LR1}.

  Let $M$ be a  differentiable manifold of dimension $n$. It is
possible to introduce an equivalence relation in the set $C^{l}(\R,
M)$ of $l$-differentiable curves from $\R$ to $M$. By definition,
two given curves in $M$, $\gamma_1(t)$ and $\gamma_2(t)$, where
$t\in (-a, a)$ with $a\in \R$ have contact of order $l$ at $q_0 =
\gamma_1(0) = \gamma_2(0)$ if there is a local chart $(\varphi, U)$
of $M$ such that $q_0 \in U$ and
\[
\frac{d^s}{dt^s}\left(\varphi \circ \gamma_1(t)\right){\Big{|}}_{t=0} =
\frac{d^s}{dt^s} \left(\varphi
\circ\gamma_2(t)\right){\Big{|}}_{t=0}\; ,
\]
for all $s = 0,...,l.$ This is a well-defined equivalence relation
in $C^{l}(\R,M)$ and the equivalence class of a  curve $\gamma$ will
be denoted by $[\gamma ]_0^{(l)}.$ The set of equivalence classes
will be denoted by $T^{(l)}M$ and it is not hard to show that it has
a natural structure of differentiable manifold. Moreover, $ \tau_M^l
: T^{(l)} M \rightarrow M$ where $\tau_M^l
\left([\gamma]_0^{(l)}\right) = \gamma(0)$ is a fiber bundle called
the \emph{tangent bundle of order $l$} of $M.$

Define the left- and right-translation of $G$ on itself
\begin{eqnarray*}
\ell:G\times G&\to&G\, ,\qquad \; (g, h)\,\mapsto \ell_g(h)=gh,\\
r:G\times G&\to&G\, ,\qquad\; (g, h)\,\mapsto r_g(h)=hg.
\end{eqnarray*} Obviously $\ell_g$ and $r_{g}$ are diffeomorphisms.

The left-translation allows us to trivialize the tangent bundle $TG$
and the cotangent bundle $T^*G$ as follows
\begin{eqnarray*}
TG&\to&G\times {\mathfrak g}, \qquad \,\,(g, \dot{g})\,\,\,\longmapsto (g, g^{-1}\dot{g})=(g, T_e\ell_{g^{-1}}\dot g)=(g, \xi),\\
T^*G&\to&G\times{\mathfrak g}^*,\qquad (g, \alpha_g)\longmapsto (g, T^*_e\ell_g(\alpha_g))=(g, \alpha),
\end{eqnarray*}
where ${\mathfrak g}=T_eG$ is the Lie algebra of $G$ and $e$ is the
neutral element of $G$. In the same way, we have the identification
$TTG\equiv G\times 3{\mathfrak g}$ where $3\al$ stands for
$\al\times\al\times\al$. Throughout this paper, the notation $n\,V$,
where $V$ is a given space, denotes the cartesian product of $n$
copies of $V$. Therefore, in the case when the manifold $M$ has a
Lie group structure, i.e. $M=G$, we can use the left trivialization
to identify the higher-order tangent bundle $T^{(l)} G$ with
$G\times l{\mathfrak g}$. That is, if $g: I\rightarrow G$ is a curve
in $C^{l}(\R, G)$, where $I\subset\R$, then:
\[
\begin{array}{rrcl}
\Upsilon^{(l)}:& T^{(l)}G&\longrightarrow& G\times l{\mathfrak g}\\
       & [g]_0^{(l)}&\longmapsto&\left(g(0), g^{-1}(0)\dot{g}(0), \frac{d}{dt}\Big|_{t=0}(g^{-1}(t)\dot{g}(t)), \ldots, \frac{d^{l-1}}{dt^{l-1}}\Big|_{t=0}(g^{-1}(t)\dot{g}(t))\right).
       \end{array}
\]
It is clear that $\Upsilon^{(l)}$ is a diffeomorphism.

We will denote  $\xi(t):=g^{-1}(t)\dot{g}(t)$, therefore
\[
\Upsilon^{(l)}([g]_0^{(l)})=(g, \xi, \dot{\xi}, \ldots, \xi^{(l-1)})\; ,
\]
where
\[
\xi^{(j)}(t)=\frac{d^{j}}{dt^{j}}(g^{-1}(t)\dot{g}(t)), \qquad 0\leq j\leq l-1
\]
and $g(0)=g, \xi^{(j)}(0)=\xi^{(j)}, 0\leq j\leq l-1$. We will use
the following notations without distinction  $\xi^{(0)}=\xi$,
$\xi^{(1)}=\dot{\xi}$, $\xi^{(2)}=\ddot{\xi}$ and so on, when
referring to the derivatives.

We   may also define the surjective mappings $\tau_G^{(j,l)} :
T^{(l)} G \rightarrow T^{(j)} G,$ for $j\leq l$, given by
$\tau_G^{(j,l)}\left([g]_0^{(l)}\right) = [g]_0^{(j)}.$ With the
previous identifications we have that
\[
\tau_G^{(j,l)} (g(0), \xi(0), \dot{\xi}(0), \ldots, \xi^{(l-1)}(0))=(g(0), \xi(0), \dot{\xi}(0), \ldots, \xi^{(j-1)}(0)).
\]
It is easy to see that $T^{(1)} G \equiv G\times {\mathfrak g}$,
$T^{(0)} G \equiv G$ and $\tau_G^{(0,l)}=\tau_G^l$.

\subsection{Euler-Lagrange equations for trivial principal
bundles} 
In this subsection and the two subsequent
ones we derive, from a variational point of view, the Euler-Lagrange
equations for the trivial principal bundle $M=Q\times G$ where $Q$
is a $n-$dimensional differentiable manifold and $G$ is a Lie group.
 We consider the cases of the tangent bundle (see
\cite{CeMaRa} for more details), second-order tangent bundle and
higher-order tangent bundle where we show the process step by step
since it may be enlightening for the reader.

Let $L:TM\ra\R$ be a Lagrangian function. Since $TG$ can be
identified with $G\times\mathfrak{g}$ after a left-trivialization,
we can consider a Lagrangian function as $L:TQ\times
G\times\mathfrak{g}\ra\R$.

The motion of the mechanical system is described by applying the
following variational principle:
\begin{equation}\label{Hprinciple}
\delta\mathcal{A}(c):=\delta\int_{0}^{T}L(c(t),\dot{c}(t))dt:=\delta\int_{0}^{T}L(q(t),\dot{q}(t),g(t),\xi(t))dt=0,
\end{equation} where $c$ is a smooth curve on $Q\times G$ and, with some abuse of notation, $\dot{c}(t)\in TQ\times G\times\mathfrak{g}$, which locally reads $\dot{c}(t)=(q(t),\dot{q}(t),g(t),\xi(t))$. The variations $\delta q(t)$ satisfy $\delta q(0)=\delta q(T)=0,$ and $\delta\xi$ verify $\delta
\xi(t)=\dot{\eta}(t)+[\xi(t),
\eta(t)]=\dot{\eta}(t)+\mbox{ad}_{\xi(t)}\eta(t)$, where $\eta(t)$
is an arbitrary curve on the Lie algebra with $\eta(0)=\eta(T)=0$
given by $\eta=g^{-1}\delta g$ (see \cite{holm}). This variational
principle gives rise to the Euler-Lagrange equations on trivial
principal bundles
\begin{subequations}\label{E-Leq1}
\begin{align}
\frac{d}{dt}\left(\frac{\partial L}{\partial\dot{q}}\right)&=\frac{\partial L}{\partial q},\label{E-Leq1a}\\
\frac{d}{dt}\left( \frac{\der L}{\der
\xi}\right)&=\hbox{ad}^*_{\xi}\left( \frac{\der L}{\der
\xi}\right)+\ell_g^*\frac{\der L}{\der g},\label{E-Leq1b}
\end{align}
\end{subequations}
where $\hbox{ad}^{*}:\mathfrak{g}^{*}\rightarrow\mathfrak{g}^{*}$ is
the coadjoint representation of the Lie algebra $\mathfrak{g}.$ If
the Lagrangian $L$ is left-invariant, that is, $L$ does not depend
on the variable on $G,$ we can perform a reduction
procedure yielding a reduced Lagrangian function $L_{\tiny
\mbox{red}}:TQ\times\mathfrak{g}\to\R$. Then the above equations
are rewritten as
\begin{subequations}\label{E-Leq2}
 \begin{align}
\frac{d}{dt}\left(\frac{\partial L_{\tiny \mbox{red}}}{\partial\dot{q}}\right)&=\frac{\partial L_{\tiny \mbox{red}}}{\partial q}\label{E-Leq2a}\\
\frac{d}{dt}\left( \frac{\der L_{\tiny \mbox{red}}}{\der
\xi}\right)&=\hbox{ad}^*_{\xi}\left( \frac{\der L_{\tiny \mbox{red}}}{\der
\xi}\right),\label{E-Leq2b}
\end{align}
\end{subequations}
which are called Lagrange-Poincar\'e equations (see \cite{CeMaRa}).

\subsection{Second-order Euler-Lagrange equations for trivial
principal bundles} 

In this subsection we develop
from a variational point of view, Euler-Lagrange equations for
second-order Lagrangian systems defined in a trivial principal
bundle, moreover, when the second-order Lagrangian is $G$-invariant
we obtain second-order Lagrange-Poincar\'e equations. These results
also appear in \cite{GHR2011} with the motivation of future studies
in computational anatomy.

Let $L: T^{(2)}Q\times G\times 2\mathfrak{g}\rightarrow\mathbb{R}$
be a Lagrangian function. The problem consists in finding the
critical curves of the action defined by
\[
\mathcal{A}(c):=\int_0^TL(c(t),\dot{c}(t),\ddot{c}(t))dt:=\int_{0}^{T}L(q(t), \dot{q}(t), \ddot{q}(t), g(t),\xi(t),\dot{\xi}(t))dt
\]
among all the smooth curves $c$ in $(Q\times G)$ with fixed endpoint
conditions. As in the previous subsection we employ the notation
$\ddot{c}(t)\in T^{(2)}Q\times G\times 2\mathfrak{g}$ which locally
reads $\ddot{c}(t)=(q(t), \dot{q}(t), \ddot{q}(t),
g(t),\xi(t),\dot{\xi}(t))$. In order to clarify the procedure of
taking variations in a Lie group we introduce here some notation. We
shall consider arbitrary variations of the curve $c$, i.e. $\delta
c=(\delta q,\delta q^{(1)},\delta q^{(2)},\delta
g,\delta\xi,\delta\dot{\xi})$, where $\delta
q:=\frac{d}{d\epsilon}|_{_{\epsilon=0}}q_{\epsilon}$, $\delta
q^{(l)}:=\frac{d^{l}}{dt^{l}}\delta q\,\,(\hbox{ for } l=1,2)$, and
$\delta g:=\frac{d}{d\epsilon}|_{_{\epsilon=0}}g_{\epsilon}.$ Here
$\epsilon\mapsto q_{\epsilon}$ and $\epsilon\mapsto g_{\epsilon}$
are smooth curves on $Q$ and $G$ respectively, for
$\epsilon\in(-a,a)\subset\R$, such that $q_0=q$ and $g_0=g.$ For any
$\epsilon,$ we define an element of the Lie algebra by
$\xi_{\epsilon}:=g^{-1}_{\epsilon}\dot{g}_{\epsilon}.$ Its
corresponding variation $\delta\xi$ induced by $\delta g$ is
$\delta\xi=\dot{\eta}+[\xi,\eta]$ where $\eta:=g^{-1}\delta
g\in\mathfrak{g}$. Therefore
\begin{eqnarray*}
\delta\mathcal{A}(c)&=&\delta\int_{0}^{T}L(q(t),\dot{q}(t),\ddot{q}(t),g(t),\xi(t),\dot{\xi}(t))dt\\
&=&\frac{d}{d\epsilon}\Big|_{\epsilon=0}\int_{0}^{T}L(q_{\epsilon}(t),\dot{q}_{\epsilon}(t),\ddot{q}_{\epsilon}(t),g_{\epsilon}(t),\xi_{\epsilon}(t),\dot{\xi}_{\epsilon}(t))dt\\
&=&\int_{0}^{T}\left(\Big\langle \frac{\partial L}{\partial \ddot{q}},\frac{d^2}{dt^2}(\delta q)\Big\rangle+\Big\langle \frac{\partial L}{\partial \dot{q}},\frac{d}{dt}(\delta q)\Big\rangle+\Big\langle \frac{\partial L}{\partial q},\delta q\Big\rangle\right.\\
&&\left.+\Big\langle \frac{\partial L}{\partial g},\delta g\Big\rangle+\Big\langle\frac{\der L}{\der\dot{\xi}},\frac{d}{dt}(\delta\xi)\Big\rangle+\Big\langle\frac{\der L}{\der\xi},\delta\xi\Big\rangle \right)dt.\\
\end{eqnarray*}
Using twice integration by parts and the endpoint
conditions $q(0)=q(T)=\dot q(0)=\dot q(T)=0$ and
$\eta(0)=\eta(T)=\dot{\eta}(0)=\dot{\eta}(T)=0,$ the stationary
condition $\delta\mathcal{A}(c)=0$ implies
\begin{eqnarray*}
0&=&\int_{0}^{T}\Big\langle\frac{d}{dt}\left(\frac{d}{dt}\frac{\partial L}{\partial\ddot{q}}-\frac{\partial L}{\partial\dot{q}}\right)+\frac{\partial L}{\partial q}
,\delta q
\Big\rangle dt\\
&&+\int_{0}^{T}\Big\langle \ell_{g}^{*}\left(\frac{\partial L}{\partial
g}\right),\eta\Big\rangle dt+\int_{0}^{T}\Big\langle\left(-\frac{d}{dt}+\ad^{*}_{\xi}\right)\left(\frac{\der
L}{\der\xi}-\frac{d}{dt}\frac{\der
L}{\der\dot{\xi}}\right),\eta\Big\rangle dt
\end{eqnarray*}
Therefore, $\delta\mathcal{A}(c)=0$ if and only if
$c\in\mathcal{C}^{\infty}(Q\times G)$ is a solution of the
second-order Euler-Lagrange equations for $L:T^{(2)}Q\times G\times
2\mathfrak{g}\ra\R,$

\begin{subequations}\label{trivial}
\begin{align}
\label{trivialb}\frac{d}{dt}\left(\frac{\partial L}{\partial\dot{q}}-\frac{d}{dt}\frac{\partial L}{\partial\ddot{q}}\right)&=\frac{\partial L}{\partial q},\\
\left(\frac{d}{dt}-\ad^{*}_{\xi}\right)\left(\frac{\der
L}{\der\xi}-\frac{d}{dt}\frac{\der
L}{\der\dot{\xi}}\right)&=\ell_{g}^{*}\frac{\partial L}{\partial g},\label{triviala}
\end{align}
\end{subequations}
which split into a $Q$ part \eqref{trivialb} and a $G$ part
\eqref{triviala}. The previous development, reaching equations
\eqref{trivial}, shall be considered as the proof of the following
result,

\begin{theorem}\label{th1} 
Let $L:T^{(2)}Q\times G\times
2\mathfrak{g}\ra\R$ be a Lagrangian where the left-trivialization
$\xi(t):=g^{-1}(t)\dot{g}(t)\in\mathfrak{g}$ has been considered,
and $\eta(t)$ is a curve on $\mathfrak{g}$ with fixed endpoints
$\eta(0)=\eta(T)=0.$ The curve $c\in\mathcal{C}^{\infty}(Q\times G)$
satisfies $\delta\mathcal{A}(c)=0$ for the action
$\mathcal{A}:\mathcal{C}^{\infty}(Q\times G)\ra\R$ given by
\[
\mathcal{A}(c)=\int_{0}^{T}L(q,\dot{q},\ddot{q},g,\xi,\dot{\xi})dt,
\]
with endpoint conditions $\delta q(0)=\delta q(T)=0$ and $\delta\dot
q(0)=\delta\dot q(T)=0$; if and only if $c$ is a solution of the
second-order Euler-Lagrange equations for $L$,
{\rm\begin{eqnarray*} \frac{\partial L}{\partial
q}-\frac{d}{dt}\frac{\partial
L}{\partial\dot{q}}+\frac{d^2}{dt^2}\frac{\partial
L}{\partial\ddot{q}}&=&0,\\
\ell_{g}^{*}\frac{\partial
L}{\partial g}+\mbox{\ad}^{*}_{\xi}\frac{\der
L}{\der\xi}-\mbox{ad}^{*}_{\xi}\left(\frac{d}{dt}\frac{\der
L}{\der\dot{\xi}}\right)-\frac{d}{dt}\frac{\der
L}{\der\xi}+\frac{d^2}{dt^2}\frac{\der
L}{\der\dot{\xi}}&=&0.
\end{eqnarray*}}
\end{theorem}

\begin{corollary}\label{Coro1}
If the Lagrangian $L:T^{(2)}Q\times G\times 2\mathfrak{g}\to\R$ is
left-invariant, that is if $L$ does not depend on $g\in G$, we can
induce a reduced Lagrangian $L_{\tiny \mbox{red}}:T^{(2)}Q\times
2\mathfrak{g}\to\R$ whose equations of motion are
{\rm\begin{subequations}\label{EP2}
\begin{align}
\frac{\partial L_{\tiny \mbox{red}}}{\partial q}-\frac{d}{dt}\frac{\partial L_{\tiny \mbox{red}}}{\partial\dot{q}}+\frac{d^2}{dt^2}\frac{\partial L_{\tiny \mbox{red}}}{\partial\ddot{q}}&=0,\label{EP2b}\\
\ad^{*}_{\xi}\frac{\der
L_{\tiny \mbox{red}}}{\der\xi}-\ad^{*}_{\xi}\left(\frac{d}{dt}\frac{\der
L_{\tiny \mbox{red}}}{\der\dot{\xi}}\right)-\frac{d}{dt}\frac{\der
L_{\tiny \mbox{red}}}{\der\xi}+\frac{d^2}{dt^2}\frac{\der
L_{\tiny \mbox{red}}}{\der\dot{\xi}}&=0.\label{EP2a}
\end{align}
\end{subequations}}
These equations are called second-order Lagrange-Poincar\'e
equations.
\end{corollary}

\subsection{Higher-order Euler-Lagrange equations on trivial principal
bundles} The previous ideas can be extended to Lagrangians defined
on a higher-order trivial principal bundle. We identify the
higher-order tangent bundle $T^{(l)}M$, for $l>2$, with
$T^{(l)}Q\times G\times l\mathfrak{g}$ after a left trivialization.

Let $L$ be Lagrangian defined on $T^{(l)}Q\times G\times
l\mathfrak{g}$, where we have local coordinates
$(q,\dot{q},\ddot{q},\ldots,q^{(l)},g,\xi,\dot{\xi},\ldots,\xi^{(l-1)}),$
$\xi=g^{-1}\dot{g}.$ Let us denote the variations
\[
\delta q=\frac{d}{d\epsilon}\Big{|}_{\epsilon=0}q_{\epsilon},\quad \delta
q^{(j)}=\left(\frac{d^{j}}{dt^j}\right)\delta q,\quad
\delta\xi^{(s)}=\frac{d^{s}}{dt^{s}}\left(\delta\xi\right),\quad
\delta g=\frac{d}{d\epsilon}\Big{|}_{\epsilon=0}g_{\epsilon}
\]
for $j=1,\ldots,l;$ $s=1,\ldots, l-1$ and $\epsilon\mapsto q_{\epsilon}$ and
$\epsilon\mapsto q_{\epsilon}$ denotes smooth curves on $Q$ and $G$
respectively with $q_{0}=q$ and $g_0=g$. The variation $\delta\xi$
is induced by $\delta g$ through $\xi=g^{-1}\dot g$ as
$\delta\xi=\dot{\eta}+[\xi,\eta]$, where $\eta$ is
the curve on the Lie algebra given by $\eta=g^{-1}\delta g$ with
fixed endpoints. Therefore, from Hamilton's principle, integrating
$k$ times by parts and using the boundary conditions
\begin{equation}\label{BoundCon}
\left\{
\begin{array}{ll}
\delta q^{(j)}(0)=\delta q^{(j)}(T)=0,\quad j=1,\ldots,l-1;&\\
\eta(0)=\dot{\eta}(0)=\ldots=\eta^{(l-1)}(0)=0,&\\
\eta(T)=\dot{\eta}(T)=\ldots=\eta^{(l-1)}(T)=0,&
\end{array}\right.
\end{equation}
(and therefore, $\delta \xi^{(s)}(0)=\delta\xi^{(s)}(T)=0,\hbox{ for
} s=1,\ldots,l-1$) we follow the same procedure as in the previous
subsections in order to obtain the higher-order Euler-Lagrange
equations for $L:T^{(l)}Q\times G\times l\mathfrak{g}\ra\R$. Define
the action functional
\begin{equation}\label{AcHO}
\mathcal{A}(c):=\int_0^TL(c^{(l)}(t))dt
\end{equation}
for $c\in\mathcal{C}^{\infty}(Q\times G)$ and $c^{(l)}(t)\in
T^{(l)}Q\times G\times l\mathfrak{g}$ locally given by
$$c^{(l)}(t)=(q(t),\dot{q}(t),\ldots,q^{(l)}(t),g(t),\xi(t),\ldots,\xi^{(l-1)}(t)),$$
and look for its critical points. The result is enclosed in the
following theorem.
\begin{theorem}\label{ContHOTPB}
Let $L:T^{(l)}Q\times G\times l\mathfrak{g}\ra\R$ be a Lagrangian
where the left-trivialization
$\xi(t):=g^{-1}(t)\dot{g}(t)\in\mathfrak{g}$ has been considered,
and $\eta(t)$ is a curve on $\mathfrak{g}$. The curve
$c\in\mathcal{C}^{\infty}(Q\times G)$ satisfies
$\delta\mathcal{A}(c)=0$ for the action functional defined in
\eqref{AcHO} if only if $c$, taking into account the endpoint
conditions \eqref{BoundCon}, is a solution of the higher-order
Euler-Lagrange equations:
\begin{eqnarray*}
\sum_{j=0}^{l}(-1)^{j}\frac{d^{j}}{dt^{j}}\left(\frac{\partial L}{\partial q^{(j)}}\right)&=&0, \\
\left(\frac{d}{dt}-\ad_{\xi}^{*}\right)\sum_{s=0}^{l-1}(-1)^{s}\frac{d^{s}}{dt^{s}}\left(\frac{\partial L}{\partial\xi^{(s)}}\right)&=&\ell_{g}^{*}\left(\frac{\partial L}{\partial g}\right) .
\end{eqnarray*}
\end{theorem}
As in the previous cases, if the Lagrangian is left-invariant the
right-hand side of the second equation vanishes and one obtains the
higher-order Lagrange-Poincar\'e equations (after introducing the
reduced Lagrangian $L_{\tiny \mbox{red}}:T^{(l)}Q\times
l\mathfrak{g}\to\R $), which coincide with the equations given in
\cite{GHR2011} for $G$-invariant Lagrangians.




\section{Discrete higher-order Lagrange-Poincar\'e equations}\label{section4}

In this section we will derive, using discrete calculus of
variations, the discrete Euler-Lagrange equations corresponding to a
Lagrangian defined on a left-trivialized higher-order tangent bundle
to $M=Q\times G$, that is, $T^{(l)}Q\times G\times l\mathfrak{g},$
where $G$ is a finite dimensional Lie group and $\mathfrak{g}$ its
Lie algebra. First we analyze the second-order case
($l=2$) since this is just a particular instance of what we consider
as the higher-order setting, i.e. an arbitrary $l$ such that
$l\geq2$; nevertheless, we detail the derivation of the discrete
second-order Euler-Lagrange equations for convenience of the reader.
The following results are an extension of previous ideas given in
\cite{CoJiMdD} to the case of trivial principal bundles.
\subsection{Discrete second-order Euler-Lagrange equations on trivial principal bundles}\label{DiscSOEU}
  A natural discretization of the
second order tangent bundle of a manifold $M$ is given by three
copies of it (see \cite{belema} for more details). Therefore we take
$3(Q\times G)\equiv 3Q\times 3G$ as a discretization of
$T^{(2)}(Q\times G)$. Next, we develop the discrete mechanics in the
case of trivial principal bundles, with the main propose of
obtaining the discrete Euler-Lagrange equations (in analogy with
\eqref{discreteeq}).

For fixed $(q_0,g_0)$, $(q_1,g_1)$, $(q_{N-1},g_{N-1}),(q_N,g_N)\in
Q\times G $, define the space of sequences
\[
\mathcal{C}^{2(N+1)}_d:=\{(q_{(0,N)},g_{(0,N)})=(q_0,q_1,\ldots,q_N,g_0,g_1,\ldots,g_N)\in
(N+1)Q\times (N+1)G\},
\]
which is isomorphic to $(N+1)\,Q\times(N+1)\,G$.
The discrete action associated with a discrete Lagrangian
$L_d:3(Q\times G)\ra\R$ is given by
\begin{equation}\label{DiscAction}
\mathcal{A}_d(q_{(0,N)},g_{(0,N)}):=\sum_{k=0}^{N-2}L_d(q_k,q_{k+1},q_{k+2},g_k,W_k,W_{k+1}),
\end{equation}
where $W_k:=g_k^{-1}g_{k+1}\in G.$ We employ this relationship
(which is called {\it reconstruction equation}), and therefore the
triple $(g_k,W_k,W_{k+1})$ instead of $(g_k,g_{k+1},g_{k+2}).$
\vspace{0.2cm}

\paragraph{{\bf Discrete Hamilton's principle for second-order trivial principal bundles:}}

{\it Hamilton's principle establishes that the
sequence $(q_{(0,N)},g_{(0,N)})\in\mathcal{C}^{2(N+1)}_d$ is a
solution of the discrete Lagrangian system determined by
$L_d:3(Q\times G)\ra\R$ if and only if $(q_{(0,N)},g_{(0,N)})$ is a
critical point of $\mathcal{A}_d.$}
\vspace{0.2cm}

We now proceed to derive the discrete equations of motion applying
discrete Hamilton's principle. For it, we consider variations of the
discrete action sum, that is,
\begin{eqnarray}
0&=&\delta\sum_{k=0}^{N-2}L_d(q_k,q_{k+1},q_{k+2},g_k,W_k,W_{k+1})\label{supersuma}\\
&=&\sum_{k=0}^{N-2}\left(\sum_{j=1}^{3}(D_jL_d|_{k})\delta q_{k+j-1}+(D_4L_d|_{k})\delta g_k+\sum_{j=5}^{6} (D_jL_d|_{k})\delta W_{k+j-5}\right),\nonumber
\end{eqnarray} where we use the notation
$D_{s}L_{d}|_{k}:=D_{s}L_{d}(q_{k},q_{k+1},q_{k+2},g_{k},W_{k},W_{k+1})$
and $D_s$ denotes the partial derivative with respect to the $s-$th
variable. Variations of $W_{k}$ are given by
considering the Lie algebra element $\Sigma_k:=g_k^{-1}\delta
g_k\in\mathfrak{g}$. Therefore, we have that
\begin{equation}\label{variaciones}
\delta W_k:=-\Sigma_k W_k+ W_k\Sigma_{k+1},
\end{equation}
where $g_k,W_k\in G$. Note that the variations of
$W_k$ \eqref{variaciones} are not {\it general ones}, but they are
determined by the left trivialization $W_k:=g_k^{-1}g_{k+1}$. In
this sense, we could say that these variations are not {\it free}
but {\it constrained}.

From now on, we will use the following notation,
\[
\begin{split}
L_d(q_k,q_{k+1},q_{k+2},g_k,W_k,W_{k+1}):=&L_d|_{(g_k,W_k,W_{k+1})}(q_k,q_{k+1},q_{k+2})\\
=&L_d|_{(q_k,q_{k+1},q_{k+2},W_k,W_{k+1})}(g_k)\\
=&L_d|_{(q_k,q_{k+1},q_{k+2},g_k)}(W_k,W_{k+1})
\end{split}
\]
and therefore
\begin{eqnarray*}
D_{i}L_{d}|_{(g_k,W_k,W_{k+1})}(q_k,q_{k+1},q_{k+2})&=&D_{i}L_d(q_k,q_{k+1},q_{k+2},g_k,W_k,W_{k+1})\hbox{ with }i=1,2,3,\\
DL_{d}|_{(q_k,q_{k+1},q_{k+2},W_k,W_{k+1})}(g_k)&=&D_{4}L_d(q_k,q_{k+1},q_{k+2},g_k,W_k,W_{k+1}),\\
D_{1}L_{d}|_{(q_k,q_{k+1},q_{k+2},g_k)}(W_k,W_{k+1})&=&D_{5}L_d(q_k,q_{k+1},q_{k+2},g_k,W_k,W_{k+1}),\\
D_{2}L_{d}|_{(q_k,q_{k+1},q_{k+2},g_k)}(W_k,W_{k+1})&=&D_{6}L_d(q_k,q_{k+1},q_{k+2},g_k,W_k,W_{k+1}).
\end{eqnarray*}
With this notation, rearranging the sum indexes,
\eqref{supersuma} can be decomposed in the following way:
\begin{eqnarray*}
\sum_{k=0}^{N-2}\sum_{j=1}^{3}(D_jL_d|_{k})\delta q_{k+j-1}&=&\sum_{k=2}^{N-2}\left(D_1L_d|_{(g_k,W_{k},W_{k+1})}(q_k,q_{k+1},q_{k+2})\right.\\
&&\left.+D_2L_d|_{(g_{k-1},W_{k-1},W_{k})}(q_{k-1},q_{k},q_{k+1})\right.\\
&&\left.+D_3L_d|_{(g_{k-2},W_{k-2},W_{k-1})}(q_{k-2},q_{k-1},q_k)\right)\delta q_k
\end{eqnarray*}
(where $q_0,q_1,q_{N-1}$ and $q_N$ have been taken fixed points and
therefore $\delta q_0=\delta q_1=\delta q_{N-1}=\delta q_{N}=0$).

The part of \eqref{supersuma} corresponding to the
variations on $G$ is decomposed as
\begin{eqnarray*}
&&\sum_{k=0}^{N-2}(D_4L_d|_{k})\delta g_k+\sum_{k=0}^{N-2}\sum_{j=5}^{6} (D_jL_d|_{k})\delta W_{j+k-5}=\\
&&\sum_{k=0}^{N-2}\left(DL_{d}|_{(q_k,q_{k+1},q_{k+2},W_k,W_{k+1})}(g_k)\delta g_k+D_1L_{d}|_{(q_k,q_{k+1},q_{k+2},g_k)}(W_{k},W_{k+1})\,\delta\,W_k\right.\\
&&\left.+D_2L_{d}|_{(q_k,q_{k+1},q_{k+2},g_k)}(W_{k},W_{k+1})\,\delta\,W_{k+1}\right)=\\
&&\sum_{k=0}^{N-2}DL_{d}|_{(q_k,q_{k+1},q_{k+2},W_k,W_{k+1})}(g_k)\,\lp g_k\Sigma_k\rp\\
&&+\sum_{k=0}^{N-2}D_1L_{d}|_{(q_k,q_{k+1},q_{k+2},g_k)}(W_{k},W_{k+1})\,\lp -\Sigma_kW_k+W_k\Sigma_{k+1}\rp\\
&&+\sum_{k=0}^{N-2}D_2L_{d}|_{(q_k,q_{k+1},q_{k+2},g_k)}(W_{k},W_{k+1})\,\lp -\Sigma_{k+1}W_{k+1}+W_{k+1}\Sigma_{k+2}\rp=\\
&&\sum_{k=2}^{N-2}\ell_{g_{k}}^{*}DL_{d}|_{(q_k,q_{k+1},q_{k+2},W_k,W_{k+1})}(g_{k})\,\Sigma_k\\
&&+\sum_{k=2}^{N-2}\left(\ell_{W_{k-1}}^{*}D_1L_d|_{(q_{k-1},q_k,q_{k+1},g_{k-1})}(W_{k-1},W_{k})-r_{W_k}^{*}D_1L_d|_{(q_{k},q_{k+1}q_{k+2},g_k)}(W_k,W_{k+1})\right.\\
&&\left.-r_{W_k}^{*}D_2L_d|_{(q_{k-1},q_k,q_{k+1},g_{k-1})}(W_{k-1},W_k)+\ell_{W_{k-1}}^{*}D_2L_d|_{(q_{k-2},q_{k-1},q_{k},g_{k-2})}(W_{k-2},W_{k-1})\right)\,\Sigma_k,
\end{eqnarray*}
where we have used that $\delta g_k=g_k\,\Sigma_k$
and $\delta W_k=-\Sigma_kW_k+W_k\Sigma_{k+1}$. Also we have
rearranged the sum index and take into account that
$\Sigma_0=\Sigma_1=\Sigma_{N-1}=\Sigma_N=0$ since $g_0,g_1,g_{N-1}$
and $g_N$ are fixed.
From these equalities we obtain the
following theorem:

\begin{theorem}\label{theoremm}
Consider the discrete curve
$(q_{(0,N)},g_{(0,N)})\in\mathcal{C}^{2(N+1)}_d$ with fixed points
$(q_0,g_0)$, $(q_1,g_1)$, $(q_{N-1},g_{N-1})$, $(q_N,g_N)$ and
variations $\delta g_k=g_k\Sigma_k$ and $\delta
W_k=-\Sigma_kW_k+W_k\Sigma_{k+1}$, where $\Sigma_k$, $k=2,...,N-2$
are arbitrary elements of the Lie algebra $\mathfrak{g}$. Then, the
discrete curve satisfies
$\delta\mathcal{A}_d(q_{(0,N)},g_{(0,N)})=0$ for
$\mathcal{A}_d:\mathcal{C}^{2(N+1)}_d\ra\R,$ given in
\eqref{DiscAction} if and only if $(q_{(0,N)},g_{(0,N)})$ satisfies
the discrete second-order Euler-Lagrange equations for
$L_{d}:·3Q\times 3G\to\R$ given by
\begin{eqnarray*}
0&=&D_1L_d|_{(g_k,W_k,W_{k+1})}(q_k,q_{k+1},q_{k+2})+D_2L_d|_{(g_{k-1},W_{k-1},W_k)}(q_{k-1},q_{k},q_{k+1})\\
&&+D_3L_d|_{(g_{k-2},W_{k-2},W_{k-1})}(q_{k-2},q_{k+1},q_{k}),\\
0&=&\ell_{g_{k}}^{*}DL_d|_{(q_k,q_{k+1},q_{k+2},W_k,W_{k+1})}(g_{k})\\
&+&\ell_{W_{k-1}}^{*}D_1L_d|_{(q_{k-1},q_k,q_{k+1},g_{k-1})}(W_{k-1},W_{k})-r_{W_k}^{*}D_1L_d|_{(q_k,q_{k+1},q_{k+2},g_k)}(W_k,W_{k+1})\\
&-&r_{W_k}^{*}D_2L_d|_{(q_{k-1},q_k,q_{k+1},g_{k-1})}(W_{k-1},W_k)+\ell_{W_{k-1}}^{*}D_2L_d|_{(q_{k-2},q_{k-1},q_k,g_{k-2})}(W_{k-2},W_{k-1}),\\
W_k&=&g_k^{-1}g_{k+1},\hbox{ for }k=2,...,N-2.
\label{reconstruction}
\end{eqnarray*}
\end{theorem}
We recall here that the discrete Lagrangian
$L_d:3Q\times 3G\to\R$ is a function of the six variables
$(q_k,q_{k+1},q_{k+2},g_k,W_k,W_{k+1})$, nevertheless, in the
equations above and below for sake of simplicity we only display the
variables involved in the partial derivatives and reorder these
derivatives with respect to them.
\begin{corollary}\label{coro}
If the discrete Lagrangian $L_d$ is G-invariant, that is, $L_d$ does
not depend on the first variable on $G,$ we may define a reduced
discrete Lagrangian {\rm $L_d^{\tiny{\mbox{red}}}:3Q\times 2G\to\R$}
and the equations in theorem \ref{theoremm} are rewritten as
\begin{eqnarray*}
0&=&D_1L_d^{\tiny{\mbox{red}}}|_{(W_k,W_{k+1})}(q_k,q_{k+1},q_{k+2})+D_2L_d^{\tiny{\mbox{red}}}|_{(W_{k-1},W_k)}(q_{k-1},q_{k},q_{k+1})\\
&&+D_3L_d^{\tiny{\mbox{red}}}|_{(W_{k-2},W_{k-1})}(q_{k-2},q_{k+1},q_{k}),\\
0&=&\ell_{W_{k-1}}^{*}D_1L_d^{\tiny{\mbox{red}}}|_{(q_{k-1},q_k,q_{k+1})}(W_{k-1},W_{k})-r_{W_k}^{*}D_1L_d^{\tiny{\mbox{red}}}|_{(q_k,q_{k+1},q_{k+2})}(W_k,W_{k+1})\\
&-&r_{W_k}^{*}D_2L_d^{\tiny{\mbox{red}}}|_{(q_{k-1},q_k,q_{k+1})}(W_{k-1},W_k)+\ell_{W_{k-1}}^{*}D_2L_d^{\tiny{\mbox{red}}}|_{(q_{k-2},q_{k-1},q_k)}(W_{k-2},W_{k-1}),\\
W_k&=&g_k^{-1}g_{k+1} \hbox{ for }k=2,...,N-2.
\end{eqnarray*}
These equations are called \textit{discrete second-order
Lagrange-Poincar\'e equations}.
\end{corollary}

\subsection{Discrete higher-order Euler-Lagrange equations on trivial principal bundles}

 It is easy to extend the presented techniques to higher-order discrete
mechanical systems. We proceed analogously to the
second-order case showed above. Consider a system determined by a
higher-order Lagrangian $L:T^{(l)}(Q\times G)\rightarrow \R$, $l\geq
1$ defined on the left-trivialized higher-order tangent bundle
$T^{(l)}Q\times G\times l\mathfrak{g}$. The associated discrete
problem is established by replacing the left-trivialized
higher-order tangent bundle by $(l+1)$ copies of $Q\times G$.

For simplicity, we use the following notation as in \cite{belema}
and \cite{CoMdDZu2013}: if $(i,j)\in (\mathbb{N}^{*})^2$ with $
i<j$, $q_{(i,j)}$ denotes the $(j-i+1)$-tupla
$(q_i,q_{i+1},...,q_{j-1},q_j).$

Let $L_d:(l+1)(Q\times G)\rightarrow\mathbb{R}$ be a discrete
Lagrangian. For fixed initial and final conditions
$\lp(q,g)_{(0,l-1)};(q,g)_{(N-l+1,N)}\rp\in (Q\times G)^{2l}$ with
$N>2l$, which stands for
\[
\lp(q_0,g_0),(q_1,g_1),...,(q_{l-1},g_{l-1});(q_{N-l+1},g_{N-l+1}),(q_{N-l+2},g_{N-l+2}),...,(q_{N},g_{N})\rp.
\]
As in the second-order case the space of discrete sequences is
defined by 
\[
\mathcal{C}^{2(N+1)}_d:=\{(q_{(0,N)},g_{(0,N)})=(q_0,q_1,\ldots,q_N,g_0,g_1,\ldots,g_N)\in
(N+1)Q\times (N+1)G\},
\]
while the discrete action associated with a discrete Lagrangian
$L_d:(l+1)(Q\times G)\Flder\R$ is given by
$\mathcal{A}_{d}:\mathcal{C}^{2(N+1)}_d\rightarrow\mathbb{R}$:
\begin{equation}\label{DiscActionl}
\mathcal{A}_{d}(q_{(0,N)},g_{(0,N)}):=\sum_{k=0}^{N-l}L_d(q_{(k,k+l)},g_k,W_{(k,k+l-1)}),
\end{equation}
where $W_k=g_k^{-1}g_{k+1}\in G.$ We recall that
$q_{(k,k+l)}=(q_k,q_{k+1},...,q_{k+l})$ and
$W_{(k,k+l-1)}=(W_k,W_{k+1},...,q_{k+l-1})$. For instance, if $l=3$,
then the discrete Lagrangian $L_{d}:4Q\times 4G\to\R$ is given by
$\displaystyle{L_d(q_k,q_{k+1},q_{k+2},q_{k+3},
g_k,W_k,W_{k+1},W_{k+2}).}$ \vspace{0.2cm}

\paragraph{{\bf Discrete Hamilton's principle for higher-order trivial principal
bundles}}\label{HPDO22} {\it Discrete Hamilton's
principle states that the sequence
$(q_{(0,N)},g_{(0,N)})\in\mathcal{C}_d^{2(N+1)}$ is a solution of
the discrete Lagrangian system determined by $L_d:(l+1)(Q\times
G)\ra\R$ with $l\geq 1$, if and only if $(q_{(0,N)},g_{(0,N)})$ is a
critical point of $\mathcal{A}_d.$}
\vspace{0.2cm}

In the following, we proceed in an analogous way to the second-order
case in order to obtain the discrete Euler-Lagrange equations.
Namely, we take variations of the discrete action sum and take into
account that $\delta g_k=g_k\Sigma_k$ and $\delta
W_k=-\Sigma_kW_k+W_k\Sigma_{k+1}$:
\begin{eqnarray*}
&&\delta\sum_{k=0}^{N-l}L_d(q_{(k,k+l)},g_k,W_{(k,k+l-1)})=\\
&&\sum_{k=0}^{N-l}\sum_{j=1}^{l+1}D_jL_d|_{(g_k,W_{(k,k+l-1)})}(q_{(k,k+l)})\,\delta q_{k+j-1}\\
&&+\sum_{k=0}^{N-l}DL_d|_{(q_{(k,k+l)},W_{(k,k+l-1)})}(g_k)\,\delta g_k\\
&&+\sum_{k=0}^{N-l}\sum_{j=1}^{l}D_jL_d|_{(q_{(k,k+l)},g_k)}(W_{(k,k+l-1)})\,\delta W_{k+j-1}=\\
&&\sum_{k=0}^{N-l}\sum_{j=1}^{l+1}D_jL_d|_{(g_k,W_{(k,k+l-1)})}(q_{(k,k+l)})\,\delta q_{k+j-1}\\
&&+\sum_{k=0}^{N-l}DL_d|_{(q_{(k,k+l)},W_{(k,k+l-1)})}(g_k)\,\lp g_k\Sigma_k\rp\\
&&+\sum_{k=0}^{N-l}\sum_{j=1}^{l}D_jL_d|_{(q_{(k,k+l)},g_k)}(W_{(k,k+l-1)})\,\lp -\Sigma_{k+j-1}W_{k+j-1}+W_{k+j-1}\Sigma_{k+j}\rp.
\end{eqnarray*}
Using the fixed endpoint conditions and rearranging the sums we
arrive to 
\begin{eqnarray*}
&&\delta\sum_{k=0}^{N-l}L_d(q_{(k,k+l)},g_k,W_{(k,k+l-1)})=\\
&&\sum_{k=l}^{N-l}\lp\sum_{j=1}^{l+1}D_jL_d|_{(g_{k-j+1},W_{(k-j+1,k-l+j)})}(q_{(k-j+1,k-j+1+l)})\rp\,\delta q_{k}\\
&&+\sum_{k=l}^{N-l}\ell_{g_k}^*\,DL_d|_{(q_{(k,k+l)},W_{(k,k+l-1)})}(g_k)\,\Sigma_k\\
&&+\sum_{k=l}^{N-l}\left(\sum_{j=1}^{l}\left( \ell^*_{W_{k-1}}D_{j}L_d|_{(q_{(k-j,k-j+l)},g_{k-j})}(W_{(k-j,k-j+l-1)})\right.\right.\\
&&\left.\left.-r^*_{W_k}D_{j}L_d|_{(q_{(k-j+1,k-j+l+1)},g_{k-j+1})}(W_{(k-j+1,k-j+l)})\rp\rp\,\Sigma_k.
\end{eqnarray*}
From these equalities we obtain the following theorem:
\begin{theorem}\label{DiscHOPB}
Consider the discrete curve
$(q_{(0,N)},g_{(0,N)})\in\mathcal{C}^{2(N+1)}_d$ with fixed points
$\lp(q,g)_{(0,l-1)};(q,g)_{(N-l+1,N)}\rp$ where $l\geq 2$ and
$N>2l$. Consider also the variations $\delta g_k=g_k\Sigma_k$ and
$\delta W_k=-\Sigma_kW_k+W_k\Sigma_{k+1}$, where $\Sigma_k$,
$k=l,...,N-l$, are arbitrary elements of the Lie algebra
$\mathfrak{g}$. Then, the discrete curve $(q_{(0,N)},g_{(0,N)})$
satisfies $\delta\mathcal{A}_d(q_{(0,N)},g_{(0,N)})=0$ for
$\mathcal{A}_d:\mathcal{C}^{2(N+1)}_d\ra\R,$ given in
\eqref{DiscActionl} if and only if $(q_{(0,N)},g_{(0,N)})$ satisfies
the discrete higher-order Euler-Lagrange equations:
\begin{eqnarray*}
0&=&\sum_{j=1}^{l+1}D_jL_d|_{(g_{k-j+1},W_{(k-j+1,k-j+l)})}(q_{(k-l+1,k-j+1+l)}),\\
0&=&\ell_{g_k}^*\,DL_d|_{(q_{(k,k+l)},W_{(k,k+l-1)})}(g_k)\\
&&+\sum_{j=1}^{l}\left( \ell^*_{W_{k-1}}D_{j}L_d|_{(q_{(k-j,k-j+l)},g_{k-j})}(W_{(k-j,k-j+l-1)})\right.\\
&&\left.-r^*_{W_k}D_{j}L_d|_{(q_{(k-j+1,k-j+l+1)},g_{k-j+1})}(W_{(k-j+1,k-j+l)})\right),\\
W_k&=&g_k^{-1}g_{k+1}, \hbox{ with }l\leq k\leq N-l.
\end{eqnarray*}

\end{theorem}
\begin{corollary}
If the discrete Lagrangian $L_d$ is G-invariant, we may introduce
the reduced discrete Lagrangian {\rm
$L_d^{\tiny{\mbox{red}}}:(l+1)Q\times lG\to\R$} and the equations in
the previous theorem are rewritten as {\rm
\begin{eqnarray*}
0&=&\sum_{j=1}^{l+1}D_jL_d^{\tiny{\mbox{red}}}|_{(W_{(k-j+1,k+l-j)})}(q_{(k-l+1,k-j+1+l)}),\\
0&=&\sum_{j=1}^{l}\left( \ell^*_{W_{k-1}}D_{j}L_d^{\tiny{\mbox{red}}}|_{(q_{(k-j,k-j+l)})}(W_{(k-j,k-j+l-1)})\right.\\
&&\left.-r^*_{W_k}D_{j}L_d^{\tiny{\mbox{red}}}|_{(q_{(k-j+1,k-j+l+1)})}(W_{(k-j+1,k-j+l)})\right),\\
W_k&=&g_k^{-1}g_{k+1} \hbox{ with }k=l,...,N-l
\end{eqnarray*}}
 These equations are considered as the discrete higher-order
Lagrange-Poincar\'e equations.
\end{corollary}




\section{Mechanical systems with constraints
on higher-order trivial principal bundles} \label{section5}

In this section we derive, from a discretization of Hamilton's
principle and using Lagrange multipliers, an integrator for
higher-order Lagrangian systems with higher-order constraints when
the configuration space is a trivial principal bundle. Previously we
derive the continuous higher-order Euler-Lagrange equations for such
systems with higher-order constraints.

\subsection{Mechanical systems defined on higher-order trivial
principal bundles subject to higher-order
constraints:}\label{subject}

Consider the Lagrangian system determined by $L:T^{(l)}Q\times
G\times l\,\mathfrak{g}\ra\R$ with constraints given by
$\Phi^{\alpha}:T^{(l)}Q\times G\times
l\,\mathfrak{g}\rightarrow\mathbb{R},\,\, 1\leq\alpha\leq m.$
 We denote by
$\mathcal{M}$ the constraint submanifold locally defined by the
vanishing of these $m$ constraint functions. Define the action
functional
\[
\mathcal{A}(c):=\int_{0}^{T}L(c^{(l)}(t))dt,
\]
where $c$ is a smooth curve in $M=Q\times G$ and, as above, we
denote by $c^{(l)}$ a curve belonging to the higher-order tangent
bundle $T^{(l)}M$ which in local coordinates reads as
\begin{equation}\label{curvee}
c^{(l)}(t)=(q(t),\dot{q}(t),\ldots,q^{(l)}(t),g(t),\xi(t),\dot{\xi}(t),\ldots,\xi^{(l-1)}).
\end{equation}
The variational principle is given by
\begin{equation}\label{problemaconstraint}\left\{
                                       \begin{array}{ll}
                                         \min \mathcal{A}(c),  & \\ \\
                                         \hbox{ subject to } \Phi^{\alpha}(c^{(l)})=0 \hbox{ for } 1\leq\alpha\leq m, &
                                       \end{array}
                                     \right.\end{equation}
where we shall consider the boundary conditions
$q(0)=q(T)=q^{(j)}(0)=q^{(j)}(T)=0,$  $j=1,\ldots,l;$
$\eta^{(s)}(0)=\eta^{(s)}(T)=0 \hbox{ for } s=0,\ldots,l-1;$ where
$\xi=g^{-1}\dot{g}$ and $\eta(t)$ is a curve in the Lie algebra
$\mathfrak{g}$ with fixed endpoints induced by the variations
$\delta\xi=\dot{\eta}+[\xi,\eta].$

\begin{definition}
A curve $c\in\mathcal{C}^{\infty}(Q\times G)$ will be called a
solution of the higher-order variational problem with constraints if
$c$ is a critical point of the problem defined by
\eqref{problemaconstraint}.
\end{definition}

Following \cite{LeMu} we characterize the regular
solutions of the higher-order variational problem with constraints
as the Euler-Lagrange equations for an extended Lagrangian
$\widetilde{L}:T^{(l)}Q\times G\times
l\mathfrak{g}\times\R^{m}\ra\R$ defined by


\[
\widetilde{L}(c^{(l)}(t),\lambda(t))=L(c^{(l)}(t))-\lambda_{\alpha}(t)\Phi^{\alpha}(c^{(l)}(t))
\]
where $\lambda_{\alpha}:I\subset\R\Flder\R$, $\alpha=1,...,m$,
which are regarded Lagrange multiplier.

The equations of motion for $\widetilde{L}$ are
\begin{eqnarray*}
0&=&\sum_{j=0}^{l}(-1)^{j}\frac{d^{j}}{dt^{j}}\left(\frac{\partial L}{\partial q^{(j)}}-\lambda_{\alpha}\frac{\partial\Phi^{\alpha}}{\partial q^{(j)}}\right),\\
0&=&\left(\frac{d}{dt}-\ad_{\xi}^{*}\right)\sum_{s=0}^{l-1}(-1)^{s}\frac{d^s}{dt^s}\left(\frac{\partial L}{\partial\xi^{(s)}}-\lambda_{\alpha}\frac{\partial\Phi^{\alpha}}{\partial\xi^{(s)}}\right)-\ell_{g}^{*}\left(\frac{\partial L}{\partial g}-\lambda_{\alpha}\frac{\partial\Phi^{\alpha}}{\partial g}\right),\\
0&=&\Phi^{\alpha}(c^{(l)}(t)),\hbox{ for }1\leq\alpha\leq m,\\
\dot{g}&=&g\xi.
\end{eqnarray*}

If the extended Lagrangian is left-invariant (that is, $\widetilde{L}$ does not depend on
the variables on $G$) these equations are rewritten as the
higher-order Lagrange-Poincar\'e equations with higher-order
constraints.
\begin{eqnarray*}
0&=&\sum_{j=0}^{l}(-1)^{l}\frac{d^{j}}{dt^{j}}\left(\frac{\partial L_{\tiny{\mbox{red}}}}{\partial q^{(j)}}-\lambda_{\alpha}\frac{\partial\Phi^{\alpha}_{\tiny{\mbox{red}}}}{\partial q^{(j)}}\right),\\
0&=&\left(\frac{d}{dt}-\ad_{\xi}^{*}\right)\sum_{s=0}^{l-1}(-1)^{s}\frac{d^s}{dt^s}\left(\frac{\partial L_{\tiny{\mbox{red}}}}{\partial\xi^{(s)}}-\lambda_{\alpha}\frac{\partial\Phi^{\alpha}_{\tiny{\mbox{red}}}}{\partial\xi^{(s)}}\right),\\
0&=&\Phi^{\alpha}(c^{(l)}(t)),\hbox{ for }1\leq\alpha\leq m.\\
\dot{g}&=&g\xi,
\end{eqnarray*}

\subsection{Discrete variational problem with constraints on
higher-order trivial principal bundles:}\label{h-ocons}

In this subsection we will get the discretization of the last
variational principle with the purpose of obtaining the discrete
higher-order Euler-Lagrangian equations for systems subject to
discrete higher-order constraints.

Let $L_d:(l+1)(Q\times G)\Flder\R$ and
$\Phi^{\alpha}_{d}:(l+1)(Q\times G)\rightarrow\mathbb{R}$ be the
discrete Lagrangian and discrete constraints, respectively, for
$1\leq\alpha\leq m$, and denote by
$\mathcal{M}_{d}\subset(l+1)(Q\times G)$ the constraint submanifold
locally determined by the vanishing of these $m$ discrete constraint
functions. As before, we define the discrete action sum by
\[
\mathcal{A}_d(q_{(0,N)},g_{(0,N)})=\sum_{k=0}^{N-l}L_d(q_{(k,k+l)},g_k,W_{(k,k+l-1)}),
\]
where $W_k=g_k^{-1}g_{k+1}$. Therefore, we can consider the
following problem as \textit{the higher-order discrete variational
problem with constraints:}
\[\left\{
                                       \begin{array}{ll}
                                         \displaystyle{\min \mathcal{A}_{d}(q_{(0,N)},g_{(0,N)})}  & \\ \\
\hbox{subject to }\Phi_{d}^{\alpha}(q_{(0,N)},g_{(0,N)})=0,\,\mbox{for}\,\,\alpha= 1,..., m,\\
                                          \end{array}
                                     \right.\]
where  $\lp(q,g)_{(0,l-1)};(q,g)_{(N-l+1,N)}\rp$ are fixed, which
stands for $q_{(0,l-1)},\,q_{(N-l+1,N)}$, $g_{(0,l-1)}$ and
$g_{(N-l+1,N)}$.

The optimization problem posed as above is equivalent to the
unconstrained higher-order discrete variational problem defined by
the discrete extended Lagrangian $\widetilde{L}_{d}:(l+1)(Q\times
G)\times\mathbb{R}^{m}\Flder\R$,
\begin{eqnarray*}
\widetilde{L}_{d}(q_{(k,k+l)},g_k,W_{(k,k+l-1)},\lambda_{\alpha}^{k})&:=&L_{d}(q_{(k,k+l)},g_k,W_{(k,k+l-1)})\\
&+&\lambda^{k}_{\alpha}\,\Phi_{d}^{\alpha}(q_{(k,k+l)},g_k,W_{(k,k+l-1)})
\end{eqnarray*}
for  $k=0,...,N-l$ and where $\lambda_{\alpha}$
are the Lagrange multipliers, $\alpha=1,...,m$ (see \cite{LeMu} for
example). Furthermore, consider the discrete action sum
\[
\widetilde{\mathcal{A}_{d}(q_{(0,N)}, g_{(0,N)},\lambda^{(0,N-l)})}:=\sum_{k=0}^{N-l}\widetilde{L}_{d}(q_{(k,k+l)},g_k,W_{(k,k+l-1)},\lambda_{\alpha}^{k}),
\]
where $\lambda^{(0,N-k)} := (\lambda^0,...,\lambda^{N-k}),$ and each
$\lambda^{j}$ is a vector with components $\lambda_{\alpha}^{j},
1\leq\alpha\leq m$. The unconstrained variational problem is defined
as the minimization of $\widetilde{\mathcal{A}}_{d}$ where
$q_{(0,l-1)},\,g_{(0,l-1)},\,q_{(N-l+1,N)},\,g_{(N-l+1,N)}$ are
fixed, $\lambda$ are free and $k=0,...,N-l$. The critical points of
the unconstrained problem will be those satisfying
\[
\delta\widetilde{\mathcal{A}}_{d}(q_{(0,N)}, g_{(0,N)},\lambda^{(0,N-l)})=0,
\]
where now we take arbitrary variations $\delta
q_{k}$, $\delta W_{k}=\Sigma_{k}W_k+W_{k}\Sigma_{k+1}$ with
$k=0,\ldots,N-2$ for arbitrary elements $\Sigma_{k}=g_{k}\delta
g_{k}\in\mathfrak{g}$, $k=0,\ldots,N-1$.
Thus, the \textit{higher-order discrete Euler-Lagrange equations
with constraints} are
\begin{eqnarray*}
0&=&\sum_{j=1}^{l+1}\left(D_jL_d|_{(g_{k-j+1},W_{(k-j+1,k-j+l)})}(q_{(k-l+1,k-j+1+l)})\right.\\
&&\left.+\lambda_{\alpha}^{k-j+1}\,D_j\Phi_d^{\alpha}|_{(g_{k-j+1},W_{(k-j+1,k-j+l)})}(q_{(k-l+1,k-j+1+l)})\right),\\
0&=&\ell_{g_k}^*\,DL_d|_{(q_{(k,k+l)},W_{(k,k+l-1)})}(g_k)+\lambda_{\alpha}^k\,\ell_{g_k}^*\,D\Phi_d^{\alpha}|_{(q_{(k,k+l)},W_{(k,k+l-1)})}(g_k)\\
&&+\sum_{j=1}^{l}\left( \ell^*_{W_{k-1}}D_{j}L_d|_{(q_{(k-j,k-j+l)},g_{k-j})}(W_{(k-j,k-j+l-1)})\right.\\
&&\left.-r^*_{W_k}D_{j}L_d|_{(q_{(k-j+1,k-j+l+1)},g_{k-j+1})}(W_{(k-j+1,k-j+l)})\right)\\
&&+\sum_{j=1}^l\left(
\lambda_{\alpha}^{k-j}\ell^*_{W_{k-1}}D_{j}\Phi_d^{\alpha}|_{(q_{(k-j,k-j+l)},g_{k-j})}(W_{(k-j,k-j+l-1)})\right.\\
&&\left.-\lambda_{\alpha}^{k-j+1}r^*_{W_k}D_{j}\Phi_d^{\alpha}|_{(q_{(k-l+1,k-j+l+1)},g_{k-j+1})}(W_{(k-j+1,k-j+l)})\right),
\end{eqnarray*}
equations valid in the range $l\leq k\leq N-l$, and also subject to
\begin{eqnarray*}
0&=&\Phi_{d}^{\alpha}(q_{(k,k+l)},g_k,W_{(k,k+l-1)}),\\
W_k&=&g_k^{-1}g_{k+1} \hbox{ for } 0\leq k\leq N-l
\end{eqnarray*} 
Note that
these equations are an extension of the higher-order Euler-Lagrange
equations obtained in theorem \ref{DiscHOPB}, where we have added
Lagrange multipliers into the picture and therefore the constrained
problem is replaced by an unconstrained problem in a larger space
(with the Lagrange multipliers).

\begin{remark}\label{referee1}
In \cite{CoMdDZu2013} we have shown that under some
regularity conditions, the discrete system with constraints
preserves a symplectic $2$-form (see Remark 3.4 in that paper). In
this sense, the methods that we are deriving are automatically
symplectic methods. Moreover, under a group of symmetries preserving
the discrete Lagrangian and the constraints, we additionally obtain
momentum preservation. The preservation of the symplectic form and
momentum map are important properties which guarantee the
competitive qualitative and quantitative behavior of the proposed
methods and are mimicking the corresponding properties of the
continuous problem to be simulated. That is, these methods can allow
substantially more accurate simulations at lower cost for
higher-order problems with constraints.

Additionally, since the methods are automatically
symplectic, the well known backward error analysis theory (see, for
instance \cite{Hair}) shows that the discrete flow associated with a
symplectic integrator applied to a Hamiltonian system can be
interpreted as the exact continuous solution of a modified
Hamiltonian system (see \cite{JdLMdD} for the relation of
constrained problems and hamiltonian ones).This fact explain the
excellent long-time energy behavior of the proposed methods since
they are preserving this modified Hamiltonian system close to the
original one. 
\end{remark}
\section{Application to optimal control of underactuated mechanical
systems}\label{section6}

The purpose of this section is to study optimal control problems in
the case of underactuated mechanical systems, that is, a Lagrangian
control system such that the number of the control inputs is fewer
than the dimension of the configuration space (also called
``superarticulated mechanical system'' following the nomenclature
given in \cite{Ba}).

Now, we introduce briefly the optimal control
problem. Consider a mechanical system which configuration space is a
differentiable manifold $M$ and whose dynamics is determined by a
Lagrangian $L:TM\Flder R$. The control forces are modeled as a
mapping $f:TM\times U\Flder T^*M$, where for $x\in M$ we have
$f(v_x,u)\in T^*_xM$, $v_x\in T_xM$ and $u\in U$, being
$U\subset\R^r$ the control space, an open subset on $\R^{r}$
containing the $0$ and $u(t)$ the control parameter. Since we are
treating the underactuated case, it follows that $r<$ dim $M$.
Observe that this last definition also covers configuration and
velocity dependent forces such as dissipation or friction. The
motion of the system is described by applying the
Lagrange-d'Alembert principle, which requires that the solution
$x(t)\in M$, where $t\in [0,T]$, must satisfy
\begin{equation}\label{LagDal}
\delta\int_0^TL(x(t),\dot x(t))dt+\int_0^Tf(x(t),\dot x(t),u(t))\,\delta x(t)dt=0,
\end{equation}
where $(x,\dot x)$ are local coordinates on $TM$ and we consider
arbitrary variations $\delta x\in T_{x(t)}M$ with $\delta
x(0)=\delta x(T)=0$. In what follows we assume that all the systems
are controllable, that is, for any two points $x_0$ and $x_f$ in the
configuration space $M$, there exists an admissible control $u(t)$
defined in $[0,T]$ such that the system with initial condition $x_0$
reaches the point $x_f$ at time $T$ (see \cite{Bl,bullolewis} for
more details). As in the previous sections consider a trivial bundle
$M=Q\times G$ and the mechanical problem reduced to $TQ\times\al$,
that is $L$ reduces to $L_{\tiny{\mbox{red}}}:TQ\times\al\to\R$ and
$f_{\tiny{\mbox{red}}}:TQ\times\al\times U\to T^{*}M\times\al^{*}$.
Moreover and for simplicity assume that the forces
$f_{\tiny{\mbox{red}}}$ are linear, that is, 
\begin{equation}\label{cforces}
f(q,\dot{q},\xi, u)=u_a\,\mathcal{B}^a(q)\in T^*Q\times\mathfrak{g}^*,
\end{equation}
where $\mathcal{B}^a$ denotes a set of $r$ linear independent
sections of the vector bundle $\pi:T^*Q\times\mathfrak{g}^*\Flder
Q$. This set of sections can be decomposed as
$\mathcal{B}^a:=(\mu^a,\eta^a)$, such that $\mu^a(q)\in T^*_qQ$ and
$\eta^a(q)\in\mathfrak{g}^*$. Taking the reduced Lagrangian
$L_{\tiny{\mbox{red}}}:TQ\times\mathfrak{g}\rightarrow\mathbb{R}$,
applying equations \eqref{LagDal} and considering the control forces
\eqref{cforces}, we obtain the following control equations
\begin{subequations}\label{control}
\begin{align}
\frac{d}{dt}\left(\frac{\partial L_{\tiny{\mbox{red}}}}{\partial \dot q}\right)-\frac{\partial L_{\tiny{\mbox{red}}}}{\partial q}&=u_a\mu^{a}(q),\label{controla}\\
\frac{d}{dt}\left( \frac{\partial L_{\tiny{\mbox{red}}}}{\partial \xi}\right)-\ad^*_{\xi}\left( \frac{\partial L_{\tiny{\mbox{red}}}}{\partial \xi}\right)&=u_a\eta^a(q).\label{controlb}
\end{align}
\end{subequations}
In order to state the optimal control problem we
need to introduce a cost functional, thus the optimal control
problem consists on finding a trajectory $(q(t), \xi(t),u(t))$ of
the state variables and control inputs satisfying \eqref{control},
subject to initial conditions $(q(0),\dot q(0), \xi(0))$ and final
conditions $(q(T),\dot q(T), \xi(T))$, and extremizing
\begin{equation}\label{Cfunctional}
{\mathcal J}(q,\xi, u)=\int_{0}^{T} C(q(t),\dot{q}(t), \xi(t),  u(t))\, dt,
\end{equation} where $C:TQ\times\al\times U\rightarrow\R$ is the cost functional.

Our purpose is to transform this optimal control
problem into a second-order variational problem with second-order
constraints. For that, complete $\mathcal{B}^{a}$ to a basis
$\mathcal{B}^A=\{\mathcal{B}^a,\mathcal{B}^{\alpha}\}$ of sections
of the vector bundle $\pi$, where
$\mathcal{B}^{\alpha}:=(\mu^{\alpha},\eta^{\alpha})$. Here
$A=(a,\alpha)$ is set such that
$A=1,...,(\mbox{dim}\,Q+\mbox{dim}\,\mathfrak{g}))=$($\mbox{dim}M$),
$a=1,...,r$ and
$\alpha=r+1,...,(\mbox{dim}\,Q+\mbox{dim}\,\mathfrak{g})$, recalling
that $r<\,(\mbox{dim}\ Q+\mbox{dim}\,\mathfrak{g})$. Take its dual
basis $\mathcal{B}_A=\{\mathcal{B}_a,\mathcal{B}_{\alpha}\}$ of
sections of the vector bundle $\tau:TQ\times{\mathfrak g}\Flder Q$.
Thus
\begin{equation}\label{deltaRel}
\bra\mathcal{B}^A,\mathcal{B}_B\ket=\delta^A_B,
\end{equation}
where $\bra\cdot,\cdot\ket$ denotes the paring between
$TQ\times{\mathfrak g}$ and $T^*Q\times{\mathfrak g}^*$. Let denote
$\mathcal{B}_a:=\{(X_a,\chi_a)\}$ and
$\mathcal{B}_{\alpha}:=\{(X_{\alpha},\chi_{\alpha})\}$ sections of
$\tau$ and introduce indexes in the coordinates of $Q$ and
$\mathfrak{g}$, namely $q^i$, $i=1,...,\mbox{dim}\,Q$ and
$s=1,...,\mbox{dim}\,\al$. Under these considerations, taking into
account \eqref{deltaRel} we can rewritte the equations
\eqref{control} as
\begin{subequations}\label{control2}
\begin{align}
\left(\frac{d}{dt}\left(\frac{\partial L_{\tiny{\mbox{red}}}}{\partial \dot q^i}\right)-\frac{\partial L_{\tiny{\mbox{red}}}}
{\partial q^i}\right)X_a^{i}(q)+\big{\langle}\frac{d}{dt}\left(\frac{\partial L_{\tiny{\mbox{red}}}}{\partial \xi}\right)-\ad_{\xi}^{*}\,\frac{\partial L_{\tiny{\mbox{red}}}}{\partial \xi},\chi_a(q)\big{\rangle}&=u_a,\label{control2a}\\
\left(\frac{d}{dt}\left(\frac{\partial L_{\tiny{\mbox{red}}}}{\partial \dot q^{i}}\right)-\frac{\partial L_{\tiny{\mbox{red}}}}
{\partial q^{i}}\right)X_{\alpha}^{i}(q)+\big{\langle}\frac{d}{dt}\left(\frac{\partial L_{\tiny{\mbox{red}}}}{\partial \xi}\right)-\ad_{\xi}^{*}\,\frac{\partial L_{\tiny{\mbox{red}}}}{\partial \xi},\chi_{\alpha}(q)\big{\rangle}&=0,\label{control2b}
\end{align}
\end{subequations}

Note that the equation \eqref{control2a} provides
a complete expression of the control variables $u_a$ in terms of the
other variables and their first and second time derivatives, i.e.
$u_a=F_{a}(q^{i},\dot{q}^{i},\ddot{q}^{i},\xi,\dot{\xi})$:
This relationship allows us to define a
second-order Lagrangian function $\widetilde{L}:T^{(2)}Q\times
2{\mathfrak g}\to \R$ by replacing the control inputs in the cost
function, namely
\[
\widetilde{L}(q^i,\dot{q}^{i},\ddot{q}^{i},\xi, \dot{\xi})=C\left(q^{i},\dot{q}^{i},\xi,F_{a}(q^{i},\dot{q}^{i},\ddot{q}^{i},\xi,\dot{\xi})\right).
\]
On the other hand, the equation \eqref{control2b} can be
reinterpreted as a set of constraints depending on the configuration
variables and their first and second time derivatives, that is,
\begin{small}
\begin{equation*}
\Phi^{\alpha}(q^{i},\dot{q}^{i},\ddot{q}^{i},\xi,\dot{\xi})=\left(\frac{d}{dt}\left(\frac{\partial L_{\tiny{\mbox{red}}}}{\partial \dot q^{i}}\right)-\frac{\partial L_{\tiny{\mbox{red}}}}
{\partial q^{i}}\right)X_{\alpha}^{i}(q)+\big{\langle}\frac{d}{dt}\left(\frac{\partial L_{\tiny{\mbox{red}}}}{\partial \xi}\right)-\ad_{\xi}^{*}\frac{\partial L_{\tiny{\mbox{red}}}}{\partial \xi},\chi_{\alpha}(q)\big{\rangle};
\end{equation*}
\end{small}

Finally, this procedure shows how the proposed optimal control
problem is equivalent to a second-order variational problem with
second-order constraints. Consequently, we can apply the techniques
introduced above to discretize.

\begin{remark}
It is possible to extend our analysis to systems with external
forces $f^{ex}:TQ\times\al\rightarrow T^{*}Q\times\al^{*}$ given by
the following diagram
\[
\xymatrix{
TQ\times \al\ar[rr]^{f^{ex}}\ar[dr]_{\tau} & &T^*Q\times\dal\ar[dl]^{\pi}\\
             &  Q&
}
\]
just by adding the corresponding term in \eqref{LagDal}, namely:
\begin{eqnarray*}
0&=&\delta\int_0^TL(q(t),\dot q(t),\xi(t))dt\\
&+&\int_0^Tf(q(t),\dot q(t),\xi(t),u(t))\,\delta(q,\xi)(t)dt+\int_0^Tf^{ex}(q(t),\dot q(t),\xi(t))\,\delta(q,\xi)(t)dt,
\end{eqnarray*}
yielding the equations of motion
\begin{subequations}
\begin{align*}
\frac{d}{dt}\left(\frac{\partial L}{\partial \dot q}\right)-\frac{\partial L}{\partial q}&=u_a\mu^{a}(q)+\hat f^{ex}(q,\dot q,\xi),\\
\frac{d}{dt}\left( \frac{\partial L}{\partial \xi}\right)-\ad^*_{\xi}\left( \frac{\partial L}{\partial \xi}\right)&=u_a\eta^a(q)+\bar f^{ex}(q,\dot q,\xi),
\end{align*}
\end{subequations}
where we decompose the external forces as
\[
\begin{array}{rrcl}
f^{ex}:& TQ\times\al&\longrightarrow& T^*Q\times\dal\\
       &(q,\dot q,\xi)&\longmapsto&(\hat f^{ex}(q,\dot q,\xi),\bar f^{ex}(q,\dot q,\xi)).
       \end{array}
\]
\end{remark}

\begin{remark}
We have shown how to transform an optimal control problem into a
second-order variational one with second-order constraints. Thus, we
may apply directly the theory developed in \S\ref{h-ocons} to obtain
suitable variational discretizations approximating the continuous
dynamics. However, one could apply a more straight discretization
strategy when dealing with this problems; more concretely, choosing
a sequence of discrete controls $\lc u_k\rc$ equations
\eqref{control} can be discretized after setting a suitable discrete
Lagrangian $L_d:Q\times Q\times G\Flder\R$ as
\begin{equation}\label{DirectDisc}
\begin{split}
D_2L_d(q_{k-1},q_k)+D_1L_d(q_{k},q_{k+1})=&(u_a)_k\mu^a(q_k),\\
\nu_{k+1}-\Ad^*_{W_k}\nu_k=&(u_a)_k\eta^a(q_k),
\end{split}
\end{equation}
where the discrete momenta are defined by
\[
\nu_k=r^*_{W_k}\frac{\der L_d}{\der W}(W_k)
\]
with $W_{k}\in G$. Moreover, we can choose a suitable discretization
of the cost function as $C_d:Q\times Q\times G\times U\Flder\R$.
With these ingredients, the discrete optimal control problem may be
defined as the minimization of
$\sum_{k=0}^{N-1}C_d(q_k,q_{k+1},W_k,u_k)$ subject to
\eqref{DirectDisc} as optimality conditions and suitable endpoint
conditions.
\end{remark}


\subsection{Optimal control of an underactuated
vehicle}

Consider a rigid body moving in $SE(2)$ with a thruster to adjust
its pose. The configuration of this system is determined by a tuple
$(x, y, \theta, \gamma)$, where $(x, y)$ is the position of the
center of mass, $\theta$ is the orientation of the blimp with
respect to a fixed basis, and $\gamma$ the orientation of the thrust
with respect to the body basis (see \cite{bullolewis} and references
therein). Therefore, following the notation $M=Q\times G$, the
configuration manifold is  $M = \mathbb{S}^1\times SE(2)$, where
$\gamma$ is the local coordinate of $\mathbb{S}^1$ and $(x, y,
\theta)$ are the local coordinates of $SE(2)$.

The Lagrangian of the system is given by its kinetic energy
\[
L(\gamma, x, y, \theta, \dot{\gamma}, \dot{x}, \dot{y}, \dot{\theta})=\frac{1}{2}m(\dot{x}^2+\dot{y}^2)+\frac{1}{2}J_1\dot{\theta}^2+\frac{1}{2}J_2( \dot{\theta} + \dot{\gamma})^2,
\]
where $m$ is the mass of the body and $J_1,\,J_2$
are the momenta of inertia around its center of mass. The control
forces are
\begin{eqnarray*}
F^1&=& \cos(\theta+\gamma)\,dx+\sin (\theta+\gamma)\, dy - p\sin \gamma d\theta,\\
F^2&=&d\gamma;
\end{eqnarray*}
where they are applied to a point on the body with distance $p > 0$
from the center of mass, along the body $x$-axis.

The possible control forces are modeled by the
codistribution determined by
$$\mathcal{F}_{c}=\hbox{span}\{F^{1},F^{2}\}.$$ Therefore, an
admissible control force would be $F_{c}=u_{1}F^{1}+u_{2}F_{2}.$

The Lagrangian is invariant under the left (trivial) action of the
Lie group $G=SE(2)$:
\begin{eqnarray*}
\varphi_{SE(2)}:SE(2)\times \mathbb{S}^{1}\times SE(2)&\Flder& \mathbb{S}^{1}\times SE(2);\\
(h,(\gamma,g))&\mapsto&(\gamma,h\,g),
\end{eqnarray*}
where the second line stands for
\[
((a, b, \alpha), (\gamma, (x, y, \theta))\mapsto
(\gamma, (x\cos\alpha-y\sin \alpha +a, x\sin\alpha + y\cos\alpha + b, \theta+\alpha),
\]
$\gamma\in\mathbb{S}^{1}, h=(a, b, \alpha)\in SE(2)$ and $g=(x, y,
\theta)\in SE(2)$. A basis of the Lie algebra $\mathfrak{se}(2)$ of
$SE(2)$ is given by
\[
e_1=\left(
\begin{array}{ccc}
0&-1&0\\
1&0&0\\
0&0&0
\end{array}
\right),\qquad
e_2=\left(
\begin{array}{ccc}
0&0&1\\
0&0&0\\
0&0&0
\end{array}
\right) ,\qquad e_3=\left(
\begin{array}{ccc}
0&0&0\\
0&0&1\\
0&0&0
\end{array}
\right),
\]
whose elements can be identified with $\hat
e_1=(1,0,0)^T$, $\hat e_2=(0,1,0)^T$ and $\hat e_3=(0,0,1)^T$
through the isomorphism $\mathfrak{se}(2)\cong\R^3$ (see
\cite{holm}). Thus, an element $\xi \in \mathfrak{se}(2)$ is of the
form $ \xi =\xi_1\, \hat e_1+\xi_2\, \hat e_2+\xi_3\, \hat e_3.$

Taking into account the $SE(2)$ left-invariance of
this system, we may consider the quotient space
$TM/SE(2)=T(\mathbb{S}^1\times SE(2))/SE(2)$ as phase space. This
quotient space is isomorphic to the product manifold
$T\mathbb{S}^1\times\mathfrak{se}(2)$, which has vector bundle
structure over $\mathbb{S}^1$ given by
$\tau_{\mathbb{S}^1}\circ\mbox{pr}_1$, where
$\mbox{pr}_1:T\mathbb{S}^1\times\mathfrak{se}(2)\Flder
T\mathbb{S}^1$ and
$\tau_{\mathbb{S}^1}:T\mathbb{S}^1\Flder\mathbb{S}^1$ are the
canonical projections. After the bundle projection $\phi:M\Flder Q$,
in this case $\phi:\mathbb{S}^1\times SE(2)\Flder \mathbb{S}^1$, we
can name the bundle structure
$\tau_{\phi}:=\tau_{\mathbb{S}^1}\circ\mbox{pr}_1:T\mathbb{S}^1\times\mathfrak{se}(2)\Flder
\mathbb{S}^1$. A section of this vector bundle (where we will denote
the space of sections by $\Gamma(\tau_{\phi})$) is given by a pair
$(Y,g),$ where $Y\in\mathfrak{X}(\mathbb{S}^1)$ and
$g:\mathbb{S}^1\Flder\mathfrak{se}(2)$ is a smooth map. A global
basis of $\Gamma(\tau_{\phi})$ is established, using the previous
notation, by:
\[
e_1=\left(\frac{\der}{\der\gamma},0\right),\,e_2=(0,\hat e_1),\,e_3=(0,\hat e_2),\,e_4=(0,\hat e_3).
\]
Analogously, we denote the global basis of
$\Gamma(\pi_{\phi})$, where
$\pi_{\phi}:T^*\mathbb{S}^1\times\mathfrak{se}(2)^*\Flder
\mathbb{S}^1$, as
\[
e_1^*=(d\gamma,0),\,e_2^*=(0,\hat e_1^*),\,e_3^*=(0,\hat e_2^*),\,e_4^*=(0,\hat e_3^*),
\]
and where the basis of $\mathfrak{se}(2)^*$ is $\hat e_i^*$,
$i=1,2,3.$

The reduced Lagrangian function in the reduced space,
$L_{\tiny{\mbox{red}}}:T\mathbb{S}^1\times\mathfrak{se}(2)\Flder\R$,
is
\[
L_{\tiny{\mbox{red}}}(\gamma,\dot{\gamma},\xi)=\frac{1}{2}m(\xi_1^2+\xi_2^2)+\frac{J_1+J_2}{2}\xi_3^2+J_2\xi_3\dot{\gamma}+\frac{J_2}{2}\dot{\gamma}^2.
\]
According to the previous notation and
\eqref{cforces}, the linear control forces are:
\begin{equation}\label{fcon1}
f(\gamma,\dot{\gamma},\xi,u)=u_1(d\gamma,0)+u_2(0,\cos{\gamma}\,\hat e_1^*+\sin{\gamma}\,\hat e_2^*-p\,\sin{\gamma}\,\hat e_3^*).
\end{equation}
Therefore we have that $\mu^1(\gamma)=d\gamma$, $\mu^2(\gamma)=0$,
$\eta^1(\gamma)=0$ and $\eta^2(\gamma)=\cos{\gamma}\,\hat
e_1^*+\sin{\gamma}\,\hat e_2^*-p\,\sin{\gamma}\,\hat e_3^*$.

Considering the reduced Lagrangian and the control forces
$f(\gamma,\dot{\gamma},\xi,u)$ in \eqref{fcon1}, then the equations
\eqref{control} are now given by
\begin{eqnarray*}
J_2(\dot{\xi}_3+\ddot{\gamma})&=&u_1,\\
m \dot{\xi_1}&=&u_2\cos{\gamma},\\
m\dot{\xi_2}+(J_1+J_2)\xi_1\xi_3+J_2\xi_1\dot{\gamma}-m\xi_1\xi_3&=&u_2\sin{\gamma},\\
(J_1+J_2)\dot{\xi}_3+J_2\ddot{\gamma}-m\xi_2(\xi_1+\xi_3)&=&-u_2p\sin{\gamma}.
\end{eqnarray*}
Furthermore, equations \eqref{control2} reads
\begin{eqnarray*}
J_2(\dot{\xi}_3+\ddot{\gamma})&=&u_1\\
m (\cos{\gamma}\dot{\xi}_1+\sin{\gamma}(\dot{\xi}_2-\xi_1\xi_3))+(J_1+J_2)\xi_1\xi_3\sin\gamma+J_2\xi_1\dot{\gamma}\sin\gamma&=&u_2,\\
m(\cos{\gamma}(\dot{\xi}_2-\xi_1\xi_3)-\sin{\gamma}\dot{\xi}_1)+\xi_1\xi_3(J_1+J_2)\cos\gamma+J_2\xi_1\dot{\gamma}\cos{\gamma}&=&0,\\
\frac{J_1+J_2}{p}(\dot{\xi}_3+p\xi_1\xi_3)+\frac{J_2}{p}(\ddot{\gamma}+p\xi_1\dot{\gamma})+m\left(\dot{\xi}_2-\xi_1\xi_3-\frac{\xi_2\xi_1+\xi_3\xi_2}{p}\right)&=&0.
\end{eqnarray*}
The optimal control problem consists on finding a trajectory of the
state variables and control inputs satisfying the las equations,
from given initial and final conditions
$(\gamma(0),\dot{\gamma}(0),\xi(0)$),
$(\gamma(T),\dot{\gamma}(T),\xi(T))$ respectively and extremizing
the functional
\[
\mathcal{J}=\int^T_0 (\rho_1u_1^2+\rho_2u_2^2)\; dt,
\]
where the cost function is given by
$C(u_1,u_2)=\rho_1u_1^2+\rho_2u_2^2$ and $\rho_1$ and $\rho_2$ are
constants denoting weights in the cost function.

As showed in the previous subsection, this optimal control problem
is equivalent to a second-order Lagrangian problem with second-order
constraints. Such a problem consists on the extremization of the
action functional
\[
\mathcal{A}(\gamma,\xi)=\int_{0}^{T}\widetilde{L}(\gamma,\dot{\gamma},\ddot{\gamma},\xi,\dot{\xi})dt,
\]
subject to constraints
$\Phi^{\alpha}(\gamma,\dot{\gamma},\ddot{\gamma},\xi,\dot{\xi})=0$,
$\alpha=1,2$.
where  $\widetilde{L}: T^{(2)}\mathbb{S}^{1}\times
2\mathfrak{se}(2)\rightarrow\mathbb{R}$ is defined by
\[
\widetilde{L}(\gamma,\dot{\gamma},\ddot{\gamma},\xi,\dot{\xi})=C(u_1(\gamma,\dot{\gamma},\ddot{\gamma},\xi,\dot{\xi}),u_2(\gamma,\dot{\gamma},\ddot{\gamma},\xi,\dot{\xi})),
\] that is,
\begin{eqnarray}
&&\widetilde{L}(\gamma,\dot{\gamma},\ddot{\gamma},\xi,\dot{\xi})=\rho_2J_2^2(\dot{\xi}_3+\ddot{\gamma})^2+\label{tildeL}\\
&&\rho_1\left(m(\cos{\gamma}\dot{\xi}_1+\sin{\gamma}(\dot{\xi}_2-\xi_1\xi_3))+(J_1+J_2)\xi_1\xi_3\sin\gamma+J_2\xi_1\dot{\gamma}\sin\gamma\right)^{2}.\nonumber
\end{eqnarray}
and the second-order constraints are given by
\begin{subequations}\label{Phid}
\begin{align}
\Phi^{1}&=m(\cos{\gamma}(\dot{\xi}_2-\xi_1\xi_3)-\sin{\gamma}\dot{\xi}_1)+\xi_1\xi_3(J_1+J_2)\cos{\gamma}+J_2\xi_1\dot{\gamma}\cos{\gamma},\label{Phida}\\
\Phi^{2}&=\frac{J_1+J_2}{p}(\dot{\xi}_3+p\xi_1\xi_3)+\frac{J_2}{p}(\ddot{\gamma}+p\xi_1\dot{\gamma})+m\left(\dot{\xi}_2-\xi_1\xi_3-\frac{\xi_2\xi_1+\xi_3\xi_2}{p}\right).\label{Phib}
\end{align}
\end{subequations}
In the following paragraphs we treat the discretization of the
variational problem as in the subsection \ref{DiscSOEU}.
\vspace{0.2cm}

Following the prescription in Theorem \ref{theoremm} and the further
conclusion in Corollary \ref{coro}, we shall consider a discrete
Lagrangian as an approximation of the associated discrete problem.
Moreover, since we are dealing with a constrained problem, we must
include the discrete constraints in the variational procedure as
shown in $\S$\ref{h-ocons}. Therefore, the discrete Lagrangian and
constraints read: $\widetilde{L_d}: 3(\mathbb{S}^{1})\times
2SE(2)\rightarrow\mathbb{R},$
$\Phi_d^{\alpha}:3(\mathbb{S}^{1})\times
2SE(2)\rightarrow\mathbb{R},$ $\alpha=1,2$. The discrete Lagrangian
$\widetilde{L_d}$ and the discrete constraints $\Phi_d^{\alpha}$ are
chosen as:
\begin{small}
\begin{eqnarray*}
&&\widetilde{L_d}(\gamma_k,\gamma_{k+1},\gamma_{k+2},W_k,W_{k+1})=\\
&&h\widetilde{L}\left(\frac{\gamma_k+\gamma_{k+1}+\gamma_{k+2}}{3},\frac{\gamma_{k+2}-\gamma_{k}}{2h},\frac{\gamma_{k+2}-2\gamma_{k+1}+\gamma_k}{h^2},\frac{\tau^{-1}(W_k)}{h},\frac{\tau^{-1}(W_{k+1})-\tau^{-1}(W_{k})}{h^2}\right),\\\\
&&\Phi^{\alpha}_d(\gamma_k,\gamma_{k+1},\gamma_{k+2},W_k,W_{k+1})=\\
&&h\Phi^{\alpha}\left(\frac{\gamma_k+\gamma_{k+1}+\gamma_{k+2}}{3},\frac{\gamma_{k+2}-\gamma_{k}}{2h},\frac{\gamma_{k+2}-2\gamma_{k+1}+\gamma_k}{h^2},\frac{\tau^{-1}(W_k)}{h},\frac{\tau^{-1}(W_{k+1})-\tau^{-1}(W_{k})}{h^2}\right),
\end{eqnarray*}
\end{small} where $\tau:\mathfrak{g}\Flder G$ is a general retraction map (see Appendix), $\widetilde{L}$ is defined in \eqref{tildeL} and $\Phi^{\alpha}$
are defined in \eqref{Phid}. As a local diffeomorphism, the
retraction map $\tau$ allows us to relate the group elements $W_k$
with the algebra elements $\xi_k$ by $\tau(h\,\xi_k)=W_k$. Roughly
speaking, this is done because the algebra is a vector space and
therefore much more handable than the Lie group; in addition, we
stay in the space where the original continuous problem is defined.
Finally, the original configuration group elements $g_k$ are
recovered from the reconstruction equation
$g_{k+1}=g_kW_k=g_k\tau(h\,\xi_k)$ as mentioned just after the
equation \eqref{DiscAction}. Moreover,
$\gamma_k,\gamma_{k+1},\gamma_{k+2}\in \mathbb{S}^1$ while
$W_k,W_{k+1}\in SE(2).$ Note that we are taking a symmetric
approximation to $\gamma(kh)$, that is
$\frac{\gamma_k+\gamma_{k+1}+\gamma_{k+2}}{3}$. Additionaly, we are
taking the usual discretizations for the first and second
derivatives, that is
$$
\dot\gamma(kh)\simeq\frac{\gamma_{k+2}-\gamma_{k}}{2h},\quad
\ddot\gamma(kh)\simeq\frac{\gamma_{k+2}-2\gamma_{k+1}+\gamma_k}{h^2},\quad
\dot\xi(kh)\simeq\frac{\xi_{k+1}-\xi_k}{h},
$$
where $\mathfrak{se}(2)\ni\xi_k=\tau^{-1}(W_k)/h$. Taking advantage
of the retraction map, we define the discrete Lagrangian and the
discrete constraints on the Lie algebra, that is, $\widetilde{L_d}:
3(\mathbb{S}^{1})\times2\mathfrak{se}(2)\rightarrow\mathbb{R},$
$\Phi_d^{\alpha}:3(\mathbb{S}^{1})\times2\mathfrak{se}(2)\rightarrow\mathbb{R},$
$\alpha=1,2$ (with some abuse of notation, we employ the same
notation, that is $\widetilde{L_d}$ and $\Phi_d^{\alpha}$, for the
Lagrangian and constraints in both spaces). The extended Lagrangian
reads
\begin{eqnarray*}
&&\widetilde{L_d}(\gamma_k,\gamma_{k+1},\gamma_{k+2},\xi_k,\xi_{k+1})+\lambda_{\alpha}^{k}\Phi^{\alpha}_d(\gamma_k,\gamma_{k+1},\gamma_{k+2},\xi_k,\xi_{k+1})=\\
&&h\widetilde{L}\left(\frac{\gamma_k+\gamma_{k+1}+\gamma_{k+2}}{3},\frac{\gamma_{k+2}-\gamma_{k}}{2h},\frac{\gamma_{k+2}-2\gamma_{k+1}+\gamma_k}{h^2},\frac{\xi_k+\xi_{k+1}}{2},\frac{\xi_{k+1}-\xi_k}{h}\right)\\
&&+\lambda_{\alpha}^k\Phi^{\alpha}\left(\frac{\gamma_k+\gamma_{k+1}+\gamma_{k+2}}{3},\frac{\gamma_{k+2}-\gamma_{k}}{2h},\frac{\gamma_{k+2}-2\gamma_{k+1}+\gamma_k}{h^2},\frac{\xi_k+\xi_{k+1}}{2},\frac{\xi_{k+1}-\xi_k}{h}\right),
\end{eqnarray*}
where again we take symmetric approximations to $\gamma_k$ and
$\xi_k.$ Finally, as in $\S$\ref{h-ocons}, applying discrete
variational calculus we obtain the discrete Euler-Lagrange equations
with discrete constraints
\begin{subequations}\label{Discrete}
\begin{align}
0&=D_1\widetilde{L_d}|_{(\xi_k,\xi_{k+1})}(\gamma_{k},\gamma_{k+1},\gamma_{k+2})+\lambda_{\alpha}^kD_1\Phi^{\alpha}_d|_{(\xi_k,\xi_{k+1})}(\gamma_{k},\gamma_{k+1},\gamma_{k+2})\label{Discretea}\\
&+D_2\widetilde{L_d}|_{(\xi_{k-1},\xi_{k})}(\gamma_{k-1},\gamma_{k},\gamma_{k+1})+\lambda_{\alpha}^{k-1}D_2\Phi^{\alpha}_d|_{(\xi_{k-1},\xi_{k})}(\gamma_{k-1},\gamma_{k},\gamma_{k+1})\nonumber\\
&+D_3\widetilde{L_d}|_{(\xi_{k-2},\xi_{k-1})}(\gamma_{k-2},\gamma_{k-1},\gamma_{k})+\lambda_{\alpha}^{k-2}D_3\Phi^{\alpha}_d|_{(\xi_{k-2},\xi_{k-1})}(\gamma_{k-2},\gamma_{k-1},\gamma_{k}),\nonumber\\\nonumber\\
0&=\Ad^*_{\tau(h\xi_{k-1})}(d\tau^{-1}_{h\xi_{k-1}})^*\lp D_1\widetilde{L_d}|_{(\gamma_{k-1},\gamma_{k},\gamma_{k+1})}(\xi_{k-1},\xi_k)+D_2\widetilde{L_d}|_{(\gamma_{k-2},\gamma_{k-1},\gamma_{k})}(\xi_{k-2},\xi_{k-1})\rp\label{Discreteb}\\
&-(d\tau^{-1}_{h\xi_{k}})^*\lp D_1\widetilde{L_d}|_{(\gamma_{k},\gamma_{k+1},\gamma_{k+2})}(\xi_k,\xi_{k+1})+D_2\widetilde{L_d}|_{(\gamma_{k-1},\gamma_{k},\gamma_{k+1})}(\xi_{k-1},\xi_{k})\rp\nonumber\\
&+\Ad^*_{\tau(h\xi_{k-1})}(d\tau^{-1}_{h\xi_{k-1}})^*\lp \lambda_{\alpha}^{k-1}D_1\Phi_d^{\alpha}|_{(\gamma_{k-1},\gamma_{k},\gamma_{k+1})}(\xi_{k-1},\xi_k)+\lambda_{\alpha}^{k-2}D_2\Phi_d^{\alpha}|_{(\gamma_{k-2},\gamma_{k-1},\gamma_{k})}(\xi_{k-2},\xi_{k-1})\rp\nonumber\\
&-(d\tau^{-1}_{h\xi_{k}})^*\lp \lambda_{\alpha}^kD_1\Phi_d^{\alpha}|_{(\gamma_{k},\gamma_{k+1},\gamma_{k+2})}(\xi_k,\xi_{k+1})+\lambda_{\alpha}^{k-1}D_2\Phi_d^{\alpha}|_{(\gamma_{k-1},\gamma_{k},\gamma_{k+1})}(\xi_{k-1},\xi_{k})\rp,\nonumber\\\nonumber\\
&k=2,...,N-2;\,\,\,\alpha=1,2\\
0&=\Phi^{\alpha}_d(\gamma_k,\gamma_{k+1},\gamma_{k+2},\xi_k,\xi_{k+1}),\,\,\,\, k=0,...,N-2;\quad \alpha=1,2.\label{Discretec}
\end{align}
\end{subequations}
As before, we only display the variables involved in the partial derivatives. To derive
\eqref{Discreteb} the properties of the right-trivialized derivative
of the retraction map and its inverse, (see appendix, proposition
\ref{Retr}) have been used (see \cite{Rabee,JKM,Marin2}). The equations \eqref{Discrete} are the ones to be solved when we want to determine the set of unknowns that we pass to detail.

In order to obtain the complete set of unknowns, that is
$\gamma_{(0,N)}, \xi_{(0,N)}, \lambda_{\alpha}^{(0,N-2)}$, we also
have to take into account the reconstruction equation, which in this
case has the form
\begin{equation}\label{recSE2}
g_{k+1}=g_{k}\tau(h\xi_k),
\end{equation}
where $g_k\in SE(2)$.

From $\S$\ref{h-ocons}, we recall that (setting
$l=2$) $q_{(0,1)},\,g_{(0,1)},\,q_{(N-1,N)},\,g_{(N-1,N)}$ are
fixed, while all the $\lambda$ are free. This can be translated in
this example as $(\gamma_0,\gamma_1)$ and
$(\gamma_{N-1},\gamma_{N})$ are fixed in the $\mathbb{S}^1$ part,
leaving $\gamma_{(2:N-2)}$ as unknowns (i.e. $N-3$ unknowns). On the
other hand, $(g_0,g_1)$ and $(g_{N-1},g_N)$ in the $SE(2)$ part are
also fixed, which by means of \eqref{recSE2} imply that $\xi_0$ and
$\xi_{N-1}$ are fixed. Nevertheless, due to the reconstruction
discretization $g_{k+1}=g_{k}\tau(h\xi_k)$, it is clear that fixing
$\xi_k$ implies constraints in the neighboring points, in this case
$g_{k+1}$ and $g_k$. If we allow $\xi_N$, that means constraints at
the points $g_{N+1}$ and $g_N$. Since we only consider time points
up to $T=Nh$, having a constraint in the beyond-terminal
configuration $g_{N+1}$ makes no sense. Hence, to ensure that the
effect of the terminal constraint on $\xi$ is correctly accounted
for, the set of algebra points must be reduced to $\xi_{(0,N-1)}$.
Furthermore, since $\xi_0$ and $\xi_{N-1}$ are also fixed, the final
set of algebra unknowns reduces to $\xi_{(1,N-2)}$ (i.e. $3(N-2)$
unknowns, since $\mbox{dim}\,\mathfrak{se}(2)=3$).

On the other hand, the boundary condition $g(T)$ is enforced by the
relation $\tau^{-1}(g_N^{-1}g(T))=0$, which means that $g_N=g(T)$.
It is possible to translate this condition in terms of algebra
elements as
\begin{equation}\label{retralge}
\tau^{-1}\lp\tau(h\xi_{N-1})^{-1}...\tau(h\xi_0)^{-1}g_0^{-1}g(T)\rp=0.
\end{equation}
We have $2(N-1)$ extra unknowns when adding the Lagrange multipliers
$\lambda_{\alpha}^{(0,N-2)}$ (recall that, in this case
$\alpha=1,2$). Summing up, we have
 \[
 (N-3)+3(N-2)+2(N-1)
 \]
 unknowns (corresponding to $\gamma_{(2,N-2)}+\xi_{(1,N-2)}+\lambda_{\alpha}^{(0,N-2)}$) for
 \[
 (N-3)+3(N-3)+3+2(N-1)
 \]
equations (corresponding to \eqref{Discretea}$+$
\eqref{Discreteb}$+$\eqref{retralge}$+$\eqref{Discretec}).
Consequently, our discrete variational problem (which comes from the
original optimal control problem) has become a nonlinear root
finding problem. From the set $\xi_{(0,N-1)}$ we can reconstruct the
configuration trajectory by means of the reconstruction equation
\eqref{recSE2}. For computational reasons it is useful to consider
the retraction map $\tau$ as the Cayley map for $SE(2)$ instead of a
truncation of the exponential map.

We also would like to stress that derivation of
these discrete equations have a pure variational formulation and as
a consequence (see \cite{mawest} for the case of first order
systems), the integrators defined in this way are symplectic,
momentum preserving and they have a good energy behavior (see
\cite{Borda},\cite{borda2}, \cite{CoMdDZu2013}, \cite{CoMaZu2011}
and Remark \eqref{referee1}).

\subsection{Optimal Control of a Homogeneous Ball on a
Rotating Plate}

 We consider the following well-known problem (see \cite{Bl,koon,LeMu}), namely the model of a homogeneous ball on
a rotating plate. A (homogeneous) ball of radius $r>0$, mass $m$ and
inertia $mk^2$ about any axis rolls without slipping on a horizontal
table which rotates with angular velocity $\Omega$
about a vertical axis $x_3$ through one of its points Apart from
the constant gravitational force, no other external forces are
assumed to act on the sphere. Let $(x, y)$ be denote the position of
the point of contact of the sphere with the table. The configuration
space of the sphere is $M=\R^2\times SO(3)$, parametrized by
$(x,y,g),$ $g\in SO(3),$ all measured with respect to the inertial
frame. Let $\omega = (\omega_1, \omega_2, \omega_3)$ be the angular
velocity vector of the sphere measured also with respect to the
inertial frame.  The potential energy is constant, so we may put
$V=0.$

The nonholonomic constraints are given by the non-slipping condition. i.e.

$$
\dot{x}+\frac{r}{2}\mbox{tr}(\dot{g}g^{T}E_2)=-\Omega y,\quad
\dot{y}-\frac{r}{2}\mbox{tr}(\dot{g}g^{T}E_1)=\Omega x,
$$ where $\{E_1,E_2,E_3\}$ is the standard basis of $\mathfrak{so}(3)$ and $\mbox{tr}$ represents the usual trace of matrices.

The matrix $\dot{g}g^{T}$ is skew-symmetric, therefore we may write

$$\dot{g}g^{T}=\left(
                 \begin{array}{ccc}
                   0 & -\omega_3 & \omega_2 \\
                   \omega_3 & 0 & -\omega_1 \\
                   -\omega_2 & \omega_1 & 0 \\
                 \end{array}
               \right)
$$
where $(\omega_1,\omega_2,\omega_3)$ represents the angular velocity
vector of the sphere. Then, we may rewrite the constraints in the
usual form:

$$
\dot{{x}}+r\omega_2=-\Omega y,\quad
\dot{{y}}-r\omega_1=\Omega x.
$$

 In addition, since we do not consider external forces the
Lagrangian of the system corresponds with the kinetic energy
\begin{eqnarray*}
K(x,y,g,\dot{x}, \dot{y},\dot{g})=\frac{1}{2}(m\dot{{x}}^2+m\dot{{y}}^2 +
mk^2(\omega_1^2 + \omega_2^2 + \omega_3^2)).
\end{eqnarray*}

Observe that the Lagrangian is metric on $M$ which is bi-invariant
on $SO(3)$ as the ball is homogeneous.

$M=\R^2\times SO(3)$ is the total space, a
 trivial principal $SO(3)$-bundle over $\R^2$ with respect the right
 $SO(3)-$action given by $(x,y,R)\mapsto (x,y,RS)$ for all $S\in
 SO(3)$ and $(x,y,R)\in \R^{2}\times SO(3).$ The action is in the
 right side since the symmetries are material symmetries.

The bundle projection $\phi:M\to Q$, in this case $\phi:\R^2\times
SO(3)\to \R^2$, is just the canonical projection on the first
factor.  Therefore, we may consider the corresponding quotient
bundle $TM/SO(3)$ over $Q=\R^2$. We will identify the tangent bundle
to $SO(3)$ with $\mathfrak{so}(3)\times SO(3)$ by using right
translation. Note that throughout the previous exposition we have
employed the left trivialization. However, we would like to point
out that the right trivialization just implies minor changes in the
derivation of the equations of motion (see \cite{holm}). An (left)
equivalent procedure was taken into account in the previous example
when defining the reduced Lagrangian $L_{\tiny{\mbox{red}}}$.

Under this identification between $T(SO(3))$ and
$\mathfrak{so}(3)\times SO(3)$, the tangent action of $SO(3)$ on
$T(SO(3))\cong \mathfrak{so}(3)\times SO(3)$ is the trivial right
action
\begin{equation}\label{Action}
\varphi_{SO(3)}:\mathfrak{so}(3)\times SO(3))\times SO(3)\to \mathfrak{so}(3)\times SO(3);\;\;\;
((\omega,R),S)\mapsto (\omega,R\,S),
\end{equation} where $\omega\in\mathfrak{so}(3)$ and $R,\,S\in SO(3).$ Thus, the quotient bundle $TM/SO(3)$ is isomorphic to the product
manifold $T\R^2\times \mathfrak{so}(3)$, and the vector bundle
projection is $\tau_{\phi}:=\tau_{\R^2}\circ\mbox{pr}_1$, where
$\mbox{pr}_1:T\R^2\times \mathfrak{so}(3)\to T\R^2$ and
$\tau_{\R^2}:T\R^2\to \R^2$ are the canonical projections.

A section of the vector bundle $\tau_{\phi}:T\R^2\times
\mathfrak{so}(3)\to \R^2$ is a pair $(X,f)$, where $X$ is a vector
field on $\R^2$ and $f:\R^2\to \mathfrak{so}(3)$ is a smooth map.
Therefore, a global basis of sections of $T\R^2\times
\mathfrak{so}(3)\to \R^2$ is
\[
   e_1=\left(\displaystyle\frac{\partial}{\partial x},0\right),
   e_2=\left(\displaystyle\frac{\partial}{\partial y},0\right),
   e_3=(0,E_1), e_4=(0,E_2), e_5=(0,E_3).
 \]
There exists a one-to-one correspondence between the space of
sections of $\tau_{\phi}$ (we denote it by $\Gamma(\tau_{\phi})$)
and the $G$-invariant vector fields on $M$. If $\lcf\cdot, \cdot
\rcf$ is the Lie bracket on the space $\Gamma(\tau_{\phi})$, then
the only non-zero fundamental Lie brackets are
\[
 \lcf e_4,e_3\rcf=e_5,\;\;\;\lcf e_5,e_4\rcf=e_3,\;\;\; \lcf
 e_3,e_5\rcf=e_4.
 \]
Moreover, it follows that the Lagrangian function $L=K$ and the
constraints are $SO(3)$-invariant. Consequently, $L$ induces a
reduced Lagrangian function $L_{\tiny{\mbox{red}}}$ on
$TM/SO(3)\simeq T\R^2\times\mathfrak{so}(3).$ Thus, we have a
constrained system on $TM/SO(3)\simeq T\R^2\times\mathfrak{so}(3)$
and note that in this case the constraints are nonholonomic and
affine in the velocities. This kind of systems was analyzed by J.
Cort\'es {\it et al} \cite{CoLeMaMa} (in particular, this example
was carefully studied). The constraints define an affine subbundle
of the vector bundle $\tau_{\phi}$ which is modeled over the vector
subbundle $\mathcal{D}$ generated by the sections
$$\mathcal{D}=\mbox{span}\{
e_5; re_1 + e_4; re_2-e_3\}.$$
Moreover, the angular momentum of the ball about the axis $x_3$ is a
conserved quantity since the Lagrangian is invariant under rotations
about the axis $x_3$ and the infinitesimal generator for these
rotations lies in the distribution $\mathcal{D}.$ The conservation
law is written as $\omega_z=c,$ where $c$ is a constant
(equivalently $\dot{\omega}_z=0$). Then by the conservation of the angular momentum
the second-order constraints appear.

After some computations the equations of motion for this constrained
system are precisely
\begin{equation}\label{qwe}
\dot{{x}}-r\omega_{2}=-\Omega y,\quad
\dot{{y}} + r\omega_{1}=\Omega x,\quad
\dot{\omega}_{3}=0,
\end{equation} together with
$$
\ddot{{x}}+\frac{k^2\Omega}{r^2+k^2}\dot{{y}}=0,\quad
\ddot{{y}}-\frac{k^2\Omega}{r^2+k^2}\dot{{x}}=0.
$$
Now, we pass to the optimization problem. Assume full controls over
the motion of the center of the ball (the shape variables). The
controlled equations of motion are:
\begin{equation}\label{contEQ}
\ddot{x}+\frac{k^2\Omega}{r^{2}+k^{2}}\dot{y}=u_1,\quad
\ddot{y}-\frac{k^{2}\Omega}{r^{2}+k^{2}}\dot{x}=u_2,
\end{equation}
where $(u_1,u_2)\in U\subset\R^2$, subject to
\begin{equation}\label{qwe}
\omega_{2}-\frac{1}{r}\dot{x}=\frac{\Omega y}{r},\quad
\omega_{1}+\frac{1}{r}\dot{y}=\frac{\Omega x}{r},\quad
\dot{\omega}_{3}=0.
\end{equation}
Next, we consider the optimal control problem for this system
following the techniques proposed in this paper.

Let $C$ be the cost function given by
\[
C(u_1,u_2)=\frac{1}{2}\left(
u_1^{2}+u_2^2\right).\;
\]
Considering fixed initial and final endpoints $(x(0),y(0)),(x(T),y(T))$; $(\dot x(0),\dot y(0))$,\\
$(\dot x(T),\dot y(T))$ in $\R^2$ and
$\omega(0),\omega(T)\in\mathfrak{so}(3)$, we look for a curve
$(q(t),\omega(t),u(t))$, where $q(t)\in\R^2$, on the reduced space
that steers the system from $(q(0),\omega(0))$ to $(q(T),\omega(T))$
minimizing
\[
\int_{0}^{T}\frac{1}{2}\left(
u_1^{2}+u_2^2\right)dt,
\]
and subject to the constraints given by equations \eqref{qwe}. Note
that $R(0),R(T)\in SO(3)$, the initial and final configurations of
the problem, are also fixed. Its dynamics is given by the
reconstruction equation $\dot R(t)=R(t)\omega(t)$.

We define the second-order Lagrangian
$\widetilde{L}:T^{(2)}\R^{2}\times 2\mathfrak{so}(3)\ra\R$
\begin{small}
\begin{equation}\label{LagSO}
\widetilde{L}(x,y,\dot{x},\dot{y},\ddot{x},\ddot{y},\omega_1,\omega_2,\omega_3,\dot{\omega}_1,\dot{\omega}_2,\dot{\omega}_3)=
\frac{1}{2}\left(\ddot{x}+\frac{k^2\Omega}{r^2+k^2}\dot{y}\right)^{2}+\frac{1}{2}\left(\ddot{y}-\frac{k^2\Omega}{r^2+k^2}\dot{x}\right)^{2},
\end{equation}
\end{small}
from
\[
\widetilde{L}(q,\dot q,\ddot q,\omega,\dot\omega)=C(u_1(q,\dot q,\ddot q,\omega,\dot\omega),u_2(q,\dot q,\ddot q,\omega,\dot\omega)),
\]
where the relationships $u_i(q,\dot q,\ddot q,\omega,\dot\omega)$,
$i=1,2$, come from \eqref{contEQ}; subject to the constraints
$\Phi^{\alpha}:T^{(2)}\R^{2}\times 2\mathfrak{so}(3)\ra\R $,
$\alpha=1,2,3,$ $$ \Phi^1=\omega_{1}+\frac{1}{r}\dot{y}-\frac{\Omega
x}{r},\quad \Phi^2=\omega_2-\frac{1}{r}\dot{x}-\frac{\Omega
y}{r},\quad \Phi^3=\dot{\omega}_3.$$ As a constrained variational
problem with constraints, the optimal control problem is prescribed
by solving the following system of 4-order differential equations
(ODEs).
\begin{eqnarray*}
0&=&\lambda_{1}\frac{\Omega}{r}+\frac{\dot{\lambda}_2}{r}+x^{(iv)}+\frac{2k^2\Omega\dddot{y}}{r^2+k^2}-\frac{k^4\Omega^{2}\ddot{x}}{(r^2+k^2)^2}\\
0&=&\lambda_2\frac{\Omega}{r}+\frac{\dot{\lambda}_1}{r}+y^{(iv)}-\frac{2k^2\Omega\dddot{x}}{r^2+k^2}-\frac{k^4\Omega^{2}\ddot{y}}{(r^2+y^2)^2},\\
0&=&\dot{\lambda_1}+\lambda_{2}\omega_3-\lambda_3\omega_2,\\
0&=&\dot{\lambda_2}-\lambda_1\omega_3+\lambda_3\omega_1,\\
0&=&\dot{\lambda_3}+\lambda_1\omega_2-\lambda_2\omega_1,\\
0&=&\omega_{1}+\frac{1}{r}\dot{y}-\frac{\Omega x}{r},\\
0&=&\omega_2-\frac{1}{r}\dot{x}-\frac{\Omega y}{r},\\
0&=&\dot{\omega}_3.
\end{eqnarray*}
In addition, the configurations $R\in SO(3)$ are given by the
reconstruction equation $\dot R=R\omega$.

In the particular case when the angular velocity $\Omega$ depends on
the time (see \cite{Bl, KM}), the equations of motion are rewritten
as
\begin{eqnarray*}
0&=&\lambda_{1}\frac{\Omega(t)}{r}+\frac{\dot{\lambda}_2}{r}+x^{(iv)}+\frac{k^2\Omega''(t)\dot{y}}{r^{2}+k^{2}}+\frac{2k^2\Omega'(t)\ddot{y}}{r^2+k^2}+\frac{2k^2\Omega(t)\dddot{y}}{r^2+k^2},\\
&+&\frac{k^2\Omega'(t)\dddot{y}}{r^2+k^2}-\frac{k^4\Omega^{2}(t)\ddot{x}}{(r^2+k^2)^2}-\frac{2k^4\Omega'(t)\Omega(t)\dot{x}}{(r^2+k^2)^2}\\
0&=&\lambda_2\frac{\Omega(t)}{r}+\frac{\dot{\lambda}_1}{r}+y^{(iv)}-\frac{k^2\Omega''(t)\dot{x}}{r^{2}+k^{2}}-\frac{3k^2\Omega'(t)\ddot{x}}{r^2+k^2}-\frac{2k^2\Omega(t)\dddot{x}}{r^2+k^2},\\
&-&\frac{k^4\Omega^{2}(t)\ddot{y}}{(r^2+y^2)^2}-\frac{2k^{4}\Omega(t)\Omega'(t)\dot{y}}{(r^2+k^2)^2},\\
0&=&\dot{\lambda_1}+\lambda_{2}\omega_3-\lambda_3\omega_2,\\
0&=&\dot{\lambda_2}-\lambda_1\omega_3+\lambda_3\omega_1,\\
0&=&\dot{\lambda_3}+\lambda_1\omega_2-\lambda_2\omega_1,\\
0&=&\omega_{1}+\frac{1}{r}\dot{y}-\frac{\Omega(t) x}{r},\\
0&=&\omega_2-\frac{1}{r}\dot{x}-\frac{\Omega(t) y}{r},\\
0&=&\dot{\omega}_3.
\end{eqnarray*}
\vspace{0.2cm}


As in the previous example, we discretize this problem by choosing a
discrete Lagrangian $\widetilde{L_d}$ and discrete constraints
$\Phi^{\alpha}_d$. We set
$\widetilde{L_d}:3(\R^2)\times2\mathfrak{so}(3)\Flder\R$ and
$\Phi_d^{\alpha}:3(\R^2)\times2\mathfrak{so}(3)\Flder\R$,
$\alpha=1,2,3$, as
\begin{eqnarray*}
&&\widetilde{L_d}(q_k,q_{k+1},q_{k+2},\omega_k,\omega_{k+1})+\lambda_{\alpha}^{k}\Phi^{\alpha}_d(q_k,q_{k+1},q_{k+2},\omega_k,\omega_{k+1})=\\
&&h\widetilde{L}\left(\frac{q_k+q_{k+1}+q_{k+2}}{3},\frac{q_{k+2}-q_{k}}{2h},\frac{q_{k+2}-2q_{k+1}+q_k}{h^2},\frac{\omega_k+\omega_{k+1}}{2},\frac{\omega_{k+1}-\omega_k}{h}\right)\\
&&+\lambda_{\alpha}^k\Phi^{\alpha}\left(\frac{q_k+q_{k+1}+q_{k+2}}{3},\frac{q_{k+2}-q_{k}}{2h},\frac{q_{k+2}-2q_{k+1}+q_k}{h^2},\frac{\omega_k+\omega_{k+1}}{2},\frac{\omega_{k+1}-\omega_k}{h}\right),
\end{eqnarray*}

We employ the same unknowns-equations counting process than in the
previous example to find out that the number of unknowns matches the
number of equations. Therefore, our discrete variational problem
(which comes from the original optimal control problem) has become
again in a nonlinear root finding problem.  As before, for
computational reasons, it is useful to consider the retraction map
$\tau$ as the Cayley map for $SO(3)$.
\medskip

To test the behavior of our numerical integrator,
we used the output of the sophisticated algorithm FSolve of Matlab
$7.6.0$ using a shooting method with sensitive derivatives to the
boundary value problem . Now we show some simulations of our method
for $T=4$, $r=1$, $\Omega=0.3$ and $\omega_3=m=k=1$; simulations
which provide the expected behavior regarding the accuracy error and
the approximation of the ``continuous dynamics'':
\begin{center}
\begin{figure}[h!]
  \centering\includegraphics[width=5.8cm]{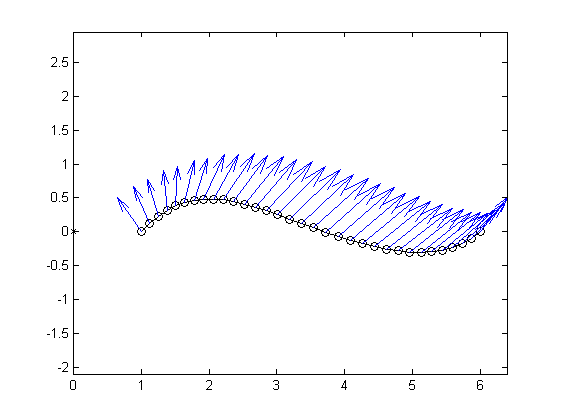}\ \includegraphics[width=7.5cm]{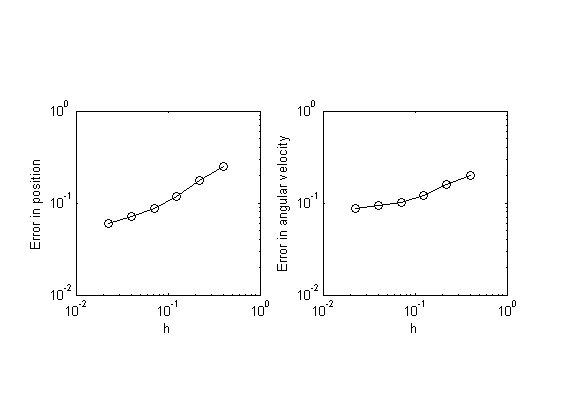}
\caption{Left: Simulation of the method with $q_0=(1,0)$
$v_0=(1,1)$, $q_N=(6,0)$, $v_N=(1,1)$, $N=33$. Blue arrows show the (scaled) angular velocity. Right: Error (root-mean-square error) in
position and angular velocity for different values of $h$.}
\end{figure}
\end{center}
The following table shows the root mean square
error in positions and angular velocities
\begin{equation*}\label{table}
\begin{tabular}{||l | c | r||} \hline \hline
h & Error in position  & Error in angular velocity \\
\hline
0.4 & 0.2471 & 0.1995\\
\hline
0.22 & 0.1746 & 0.1576 \\
\hline
0.1250&0.1173&0.1204\\
\hline
0.0714&0.0866&0.1020\\
\hline
0.04&0.0705&0.0932\\
\hline
0.0225&0.0606&0.0875\\
\hline
\end{tabular}
\end{equation*}

\begin{center}
\begin{figure}[h]
  \centering\includegraphics[width=6.5cm]{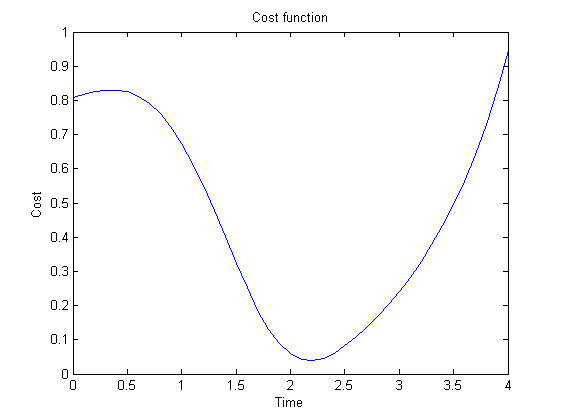}
\caption{Cost function $\frac{1}{2}(u_{1}^{2}+u_{2}^{2})$}
\end{figure}
\end{center}

\begin{center}
\begin{figure}[h]
  \centering\includegraphics[width=6.5cm]{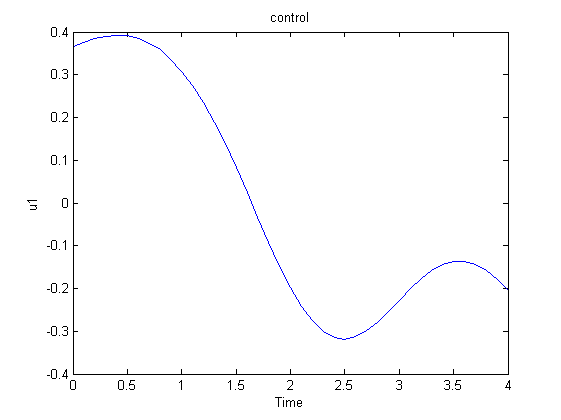}\ \includegraphics[width=6.5cm]{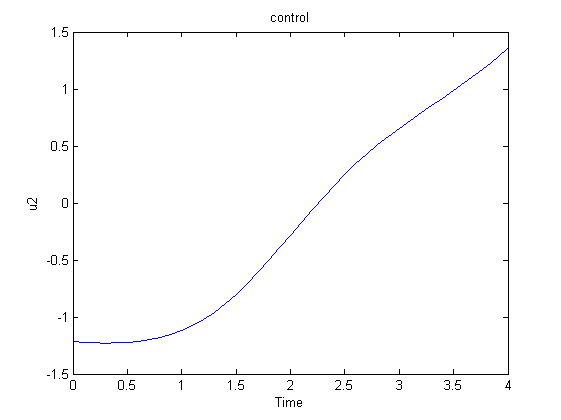}
\caption{Controls $u_1$ and $u_2$}
\end{figure}
\end{center}
Observe that directly from construction the method
is symplectic.  We expect that this family of symplectic integrators
could quantitative and qualitative benefit from structure
preservation (see \cite{chiba} and references therein).


\section*{Conclusions}

In this paper, we have designed a new class of variational
integrators for optimal control problems of underactuated mechanical
systems, showing how developments in the theory of discrete
mechanics and discrete calculus of variations with constraints can
be used to construct numerical algorithms for optimal control
problems with certain geometric desirable features,v(these features follow directly from the variational
structure and were showed in \cite{belema},
\cite{Borda},\cite{borda2}, \cite{CoMaZu2011} and
\cite{CoMdDZu2013}. Also see Remark \eqref{referee1}).

We construct variational principles for higher-order systems with
higher-order constraints (both in the continuous and discrete
settings) and we use these developments to solve optimal control
problems where the configuration space is a trivial principal
bundle. From a discretization of Hamilton's principle we derive the
discrete version of the problem. We show two concrete applications
of our ideas: the optimal control of an underactuated vehicle and a
(homogeneous) ball rotating on a plate.


It is our intention to extend this construction to the case
of non-trivial fiber bundles  using a connection to split the
reduced space \cite{CeMaRa}.


\section*{Appendix: Retraction maps}
\label{section2}

In this appendix we will review the basics notions about retraction
maps and the Cayley transformation (as an example of retraction map)
which we use along this work.

A way to  discretize a continuous problem is using a
\emph{retraction map} $\tau: {\mathfrak g}\to G$ which is an
analytic local diffeomorphism. This application maps a neighborhood
of $0\in {\mathfrak g}$ to a neighborhood of the identity $e\in G$.
As a consequence, it is possible to deduce that
$\tau(\xi)\tau(-\xi)=e$ for all $\xi \in \mathfrak{g}$.

The retraction map  is used to express small discrete changes in the
group configuration through unique Lie algebra elements (see
\cite{KM}), namely $\xi_k=\tau^{-1}(g_{k}^{-1}g_{k+1})/h$, where
$\xi_k\in\mathfrak{g}$. That is, if $\xi_{k}$ were regarded as an
average velocity between $g_{k}$ and $g_{k+1}$, then $\tau$ is an
approximation to the integral flow of the dynamics. The difference
$g_{k}^{-1}\,g_{k+1}\in G$, which is an element of a nonlinear
space, can now be represented by the vector $\xi_{k},$ in order to
enable unconstrained optimization in the linear space $\mathfrak{g}$
for optimal control purposes.

It will be useful in the sequel, mainly in the derivation of the
discrete equations of motion, to define the {\it right trivialized}
tangent retraction map as a  function
$d\tau:\mathfrak{g}\times\mathfrak{g}\rightarrow\mathfrak{g}$ by

\[
T\tau(\xi)\cdot\eta=Tr_{\tau(\xi)}d\tau_{\xi}(\eta),
\]
where $\eta\in\mathfrak{g}$. Here we use the following notation,
$d\tau_\xi:=d\tau(\xi):\mathfrak{g}\rightarrow\mathfrak{g}.$ The
function $d\tau$ is linear in its second argument. From this
definition the following identities hold (see \cite{Rabee} for
further details)

{\bf Proposition:}\label{Retr}
{\it Given a map $\tau:\mathfrak{g}\Flder G$, its right trivialized
tangent $d\tau_{\xi}:\mathfrak{g}\Flder\mathfrak{g}$ and its inverse
$d\tau_{\xi}^{-1}:\mathfrak{g}\Flder\mathfrak{g}$, are such that for
$g=\tau(\xi)\in G$ and $\eta\in\mathfrak{g}$, the following
identities hold:
\[
\der_{\xi}\tau(\xi)\,\eta=\mbox{d}\tau_{\xi}\,\eta\,\tau(\xi)\,\,\hbox{ and }\,\,
\der_{\xi}\tau^{-1}(g)\,\eta=\mbox{d}\tau^{-1}_{\xi}(\eta\,\tau(-\xi)).
\]}

The most natural example of a retraction map is the exponential map at
the identity $e$ of the group $G,$ $\e_{e}:\mathfrak{g}\rightarrow
G$. We recall that, for a finite-dimensional Lie group, $\e_e$ is
locally a diffeomorphism and gives rise to a natural chart
\cite{Mars3}. Then, there exists a neighborhood $U$ of $e\in G$ such
that $\e_e^{-1}:U\rightarrow \e_e^{-1}(U)$ is a local
$\mathcal{C}^{\infty}-$diffeomorphism. A chart at $g\in G$ is given
by $\Psi_{g}=\e_{e}^{-1}\circ \ell_{g^{-1}}.$

In general, it is not easy to work with the exponential map. In
consequence it will be useful to use a different retraction map.
More concretely, the Cayley map (see \cite{Rabee, Hair} for further
details) will provide us a proper framework in the examples shown
along the paper.

\subsection*{The Cayley map}
The Cayley map $\ca:\al\Flder G$ is defined by
\[
\ca(\xi)=(e-\frac{\xi}{2})^{-1}(e+\frac{\xi}{2})
\]
and is valid for a class of quadratic groups (see \cite{Hair} for
example) that include the groups of interest in this paper (e.g.
$SO(3)$, $SE(2)$ and $SE(3)$). Its right trivialized derivative and
inverse are defined by
$$
\mbox{d}\ca_{x}\,y=(e-\frac{x}{2})^{-1}\,y\,(e+\frac{x}{2})^{-1},\quad
\mbox{d}\ca_{x}^{-1}\,y=(e-\frac{x}{2})\,y\,(e+\frac{x}{2}).
$$

\subsubsection*{\textbf{The Cayley map for $SE(2)$:}}

The coordinates on $SE(2)$ are $(\theta, x, y)$ with matrix
representation for $g\in SE(2)$ given by
\[
g=\lp\begin{array}{ccc}
\mbox{cos}\,\theta&-\mbox{sin}\theta&x\\
\mbox{sin}\,\theta&\mbox{cos}\theta&y\\
0&0&1
\end{array}
\rp.
\]
Using the isomorphic map $\hat\cdot:\R^{3}\Flder\se$ given by
\[
\hat v=\lp\begin{array}{ccc}
0&-v_{1}&v_{2}\\
v_{1}&0&v_{3}\\
0&0&0
\end{array}
\rp,
\]
where $v=(v_{1},v_{2},v_{2})^{T}\in\R^{3}$, the set $\lc\hat
e_{1},\hat e_{2},\hat e_{3}\rc$ can be used as a basis for $\se$,
where $\lc e_{1},e_{2},e_{3}\rc$ is the standard basis of $\R^{3}$.
The map $\ca:\se\Flder SE(2)$ is given by
\begin{equation}\nonumber
\ca(\hat v)=\lp\begin{array}{c}
\frac{1}{4+v_{1}^{2}}\lp\begin{array}{ccc}
4-v_{1}^{2}&-4v_{1}&-2v_{1}v_{3}+4v_{2}\\
4v_{1}&4-v_{1}^{2}&2v_{1}v_{2}+4v_{3}
\end{array}
\rp\\
\begin{array}{ccc}
0\,\,\,\,\,\,\,\,&\,\,\,\,\,\,0\,\,\,\,\,\,&\,\,\,\,\,\,\,\,\,\,1
\end{array}
\end{array}
\rp,
\end{equation}
while the map $\mbox{d}\tau_{\xi}^{-1}$ becomes the $3\times 3$
matrix
\[
\mbox{d}\ca_{\hat v}^{-1}=I_{3}-\frac{1}{2}\ad_{v}+\frac{1}{4}\lp v_{1} v\,\,\,0_{3\times2}\rp,
\]
where
\[
\ad_{v}=\lp\begin{array}{ccc}
0&0&0\\
v_{3}&0&-v_{1}\\
-v_{2}&v_{1}&0
\end{array}
\rp
\] and $I_3$ denotes the $3\times 3$ identity matrix.

\subsubsection*{\textbf{The Cayley map for $SO(3)$:}} The group of rigid body
rotations is represented by $3\times 3$ matrices with orthonormal
column vectors corresponding to the axes of a right-handed frame
attached to the body. On the other hand, the algebra $\alg$ is the
set of $3\times 3$ antisymmetric matrices. A $\alg$ basis can be
constructed as $\lc\hat e_{1},\hat e_{2},\hat e_{3}\rc$, $\hat
e_{i}\in\alg$, where $\lc e_{1},e_{2},e_{3}\rc$ is the standard
basis for $\R^{3}$. Elements $\xi\in\alg$ can be identified with the
vector $\omega\in\R^{3}$ through $\xi=\omega^{\alpha}\,\hat
e_{\alpha}$, or $\xi=\hat\omega$. Under such identification the Lie
bracket coincides with the standard cross product, i.e.,
$\ad_{\hat\omega}\,\hat\rho=\omega\times\rho$, for some
$\rho\in\R^{3}$. Using this identification we have
\begin{equation}\label{caySO}
\ca(\hat\omega)=I_{3}+\frac{4}{4+\parallel\omega\parallel^{2}}\lp\hat\omega+\frac{\hat\omega^{2}}{2}\rp,
\end{equation}
where $I_{3}$ is the $3\times 3$ identity matrix. The linear maps
$\mbox{d}\tau_{\xi}$ and $\mbox{d}\tau_{\xi}^{-1}$ are expressed as
the $3\times 3$ matrices
\begin{equation}\label{Dtau}
\mbox{d}\ca_{\omega}=\frac{2}{4+\parallel\omega\parallel^{2}}(2I_{3}+\hat\omega),\,\,\,\,\,\mbox{d}\ca_{\omega}^{-1}=I_{3}-\frac{\hat\omega}{2}+\frac{\omega\,\omega^{T}}{4}.
\end{equation}



\end{document}